\documentclass[useAMS,usenatbib]{mn2e}

\usepackage{times}  
\usepackage{listings}
\usepackage{graphics}
\usepackage{subfigure}
\usepackage{graphicx}
\usepackage{graphicx,xspace}
\usepackage{amssymb}
\usepackage{hyperref}
\usepackage{aas_macros}
\usepackage{lscape}
\usepackage{rotating}
\usepackage{placeins}
\usepackage{natbib}
\usepackage{color}

\newcommand{\galform}{{\sc{galform}}\xspace}
\newcommand{\grasil}{{\sc{grasil}}\xspace}

\title[The stellar masses of galaxies]
{How well can we really estimate the stellar masses of galaxies from broad-band photometry?}
\author[Peter D. Mitchell]{
\parbox[t]{\textwidth}{
Peter D. Mitchell\thanks{\rm E-mail: peter.mitchell@durham.ac.uk }, 
Cedric G. Lacey,
Carlton M. Baugh,
Shaun Cole
}
\vspace*{6pt} \\
Institute for Computational Cosmology, Department of Physics,
University of Durham, South Road, Durham, DH1 3LE, UK.
\vspace*{-0.5cm}}

\begin{document}
\date{\today}
\pagerange{\pageref{firstpage}--\pageref{lastpage}} \pubyear{2013}
\maketitle
\label{firstpage}

\begin{abstract}
The estimated stellar masses of galaxies are widely used to characterize how the galaxy population 
evolves over cosmic time. 
If stellar masses can be estimated in a robust manner, free from any bias, global 
diagnostics such as the stellar mass function can be used to constrain the physics of galaxy 
formation.
We explore how galaxy stellar masses, estimated by fitting broad-band 
spectral energy distributions (SEDs) with stellar population models, can be biased as a result of commonly adopted assumptions for the 
star formation (SF) and chemical enrichment histories, recycled fractions and dust attenuation curves of 
galaxies. 
We apply the observational technique of broad-band SED fitting to model galaxy SEDs calculated by 
the theoretical galaxy formation model \galform, isolating the effect of each of these assumptions.
We find that, averaged over the entire galaxy population, the common assumption of exponentially declining
star-formation histories does not, by itself, adversely affect stellar mass estimation.
However, we also show that this result does not hold when considering galaxies that have undergone a recent
burst of star formation.
We show that fixing the metallicity in SED fitting or using sparsely sampled metallicity 
grids can introduce mass dependent systematics into stellar mass estimates. 
We find that the common assumption of a star-dust geometry corresponding to a uniform foreground dust 
screen can cause the stellar masses of dusty model galaxies to be significantly underestimated. 
Finally, we show that stellar mass functions recovered by applying SED fitting to model galaxies 
at high redshift can differ significantly in both shape and normalization from the intrinsic mass 
functions predicted by a given model. In particular, the effects of dust can reduce the normalization
at the high mass end by up to $0.6 \, \mathrm{dex}$ in some cases.
Given these differences, our methodology of using stellar masses 
estimated from model galaxy SEDs offers a new, self-consistent way to compare model predictions with observations. 
We conclude that great care should be taken when comparing theoretical galaxy formation models to 
observational results based on the estimated stellar masses of high redshift galaxies.
\end{abstract}


\section{Introduction}
\label{Introduction}

A successful theory of galaxy formation is essential for accurately connecting any underlying cosmological framework with the observable Universe. The
vast dynamic range and overall complexity involved in the interplay between gas, stars and dark matter in galaxies strongly restricts what any model of 
galaxy formation can predict \emph{a priori}. To make progress, it is necessary to use observational results to 
constrain galaxy formation models. Estimating the stellar masses of galaxies offers, in principle, a powerful method to characterize the galaxy population 
that can be compared directly to theoretical predictions. Unlike directly observable quantities, stellar mass is a derived quantity that can only be 
estimated from observational data through the application of a series of models and assumptions. It is therefore critical to understand how these assumptions affect 
the reliability of stellar mass estimates and how any uncertainties affect global diagnostics of the galaxy population, such as the stellar mass function. 

The traditional approach for constraining parameters in theoretical galaxy formation models is to use directly observable properties of the galaxy population 
such as the luminosity function or the Tully-Fisher relation. This requires that the intrinsic physical properties of model galaxies calculated using 
either hydrodynamical simulations or semi-analytic models (SAMs) can be converted into directly observable quantities. Stellar population synthesis (SPS) models \cite[e.g.][]{Bruzual03, Maraston05}
are combined with predicted star-formation and chemical enrichment histories of model galaxies to produce spectral energy distributions (SEDs) that can be
compared with observational data. Uncertainties in the form of the initial mass function (IMF) of stars, the accuracy of SPS models and the manner in which
dust attenuates the light emitted by stars make this process challenging \cite[e.g.][]{Conroy09, Conroy10a}. It can be difficult in some cases to be confident whether the comparison 
between model predictions and observational results does in fact show if a given model is successful in describing the underlying physics.

The estimation of the stellar masses of galaxies from observational data is typically achieved using the technique of broad-band SED fitting, which inverts the process 
of generating observables from intrinsic galaxy properties. The popularity of this technique can be attributed to the success of multi-wavelength 
surveys such as the Sloan Digital Sky Survey \cite[SDSS;][]{York00}, the Great Observatories Origins Deep Survey \cite[GOODS;][]{Giavalisco04} and the
Cosmological Evolution Survey \cite[COSMOS;][]{Scoville07}. These surveys quantify how the galaxy population evolves with look-back time
by utilizing broad-band photometry from the UV to the NIR to obtain accurate photometric redshifts for much larger galaxy samples than would be possible
using spectroscopy. The information and methods used to obtain accurate photometric redshifts by fitting stellar population models can readily be extended to 
also estimate stellar masses and other galaxy properties. Consequently, it has become standard practice to estimate these quantities whenever multi-wavelength 
photometry is available.

The same uncertainties associated with converting the intrinsic properties of model galaxies into observables also affect the accuracy of SED fitting applied
to observational data. 
However, there are a number of additional assumptions that must be made to estimate the stellar masses of galaxies from observations. For example, 
unlike for the case of model galaxies, the star formation histories (SFHs) of real galaxies are not known. Instead, a prior for the SFHs of galaxies must 
be adopted. Various studies  have attempted to explore how stellar mass estimates depend on the different assumptions made in SED fitting 
\cite[e.g.][]{Papovich01,Wuyts07, Marchesini09, Conroy09, Maraston10, Ilbert10, Michalowski12, Banerji13, Moustakas13, Schaerer13}. There is a general consensus that stellar 
mass estimates are more reliable compared to other quantities such as star-formation rates (SFRs) and galaxy ages that can be estimated using the same process. 
However, the reported level of uncertainty on stellar mass estimates can vary strongly, depending on the specific galaxy samples considered. 
\cite{Conroy09} find that the stellar mass-to-light ($M/L$) ratios of star-forming galaxies at $z=0$ can only be constrained to within $0.3 \, \mathrm{dex}$ at 
a $95\%$ level of confidence when various uncertainties in SPS modelling are taken into account. This does not include the uncertainty associated with the choice of the IMF. 
\cite{Gallazzi09} find that, ignoring the uncertainties associated with SPS modelling and dust attenuation, it is possible to constrain $M/L$ ratios of 
galaxies with smooth SFHs to within $0.1 \, \mathrm{dex}$ using spectral features or a single optical colour \cite[see also][]{Wilkins13}. \cite{Longhetti09} consider the accuracy 
of stellar mass estimates of early-type galaxies, finding that the true stellar masses of mock galaxies can only be recovered to within a factor of 
$\approx 0.3-0.5 \, \mathrm{dex}$, given the variations between different SPS models and metallicities. \cite{Marchesini09} quantify how uncertainties associated with the
assumptions made in SED fitting contribute to the error budget of the stellar mass function. They find that potential systematic errors associated with 
these assumptions can dominate over other error sources, such as photometric redshifts or galaxy photometry. They also find that the shape of the mass 
function, particularly at the low-mass end, is sensitive to, for example, the assumed metallicity and the adopted dust attenuation law.

We explore this topic from an alternative angle by applying the methods used in observational studies to estimate stellar masses from
SEDs output by the semi-analytic galaxy formation model \galform \citep{Cole00}. We focus on understanding the various random and systematic errors encountered when estimating the stellar 
masses of individual galaxies and also study how these translate into errors in the stellar mass function. This exercise serves as a useful example of how the process of converting 
between observables and intrinsic galaxy properties can have an impact on global diagnostics of the galaxy population. Several studies have adopted a similar approach by combining 
SED fitting with galaxy formation models, aiming to understand the accuracy of SED fitting in different scenarios \citep{Lee09, Wuyts09, Lee10, Pforr12}. This method is 
useful because outside of the limited number of cases where additional data are available, it is difficult to test the accuracy of quantities estimated using SED fitting. 
Fitting mock galaxy SEDs provides a means to do this and theoretical galaxy formation models are, in principle, a useful tool for producing samples of mock galaxies with realistic 
star-formation and chemical enrichment histories. In the case of SAMs, model galaxy samples that represent the entire galaxy population over a range of redshifts can be 
generated rapidly. This allows us to isolate and understand different effects by considering variants of the underlying model. 

\cite{Lee09} use model galaxy SEDs from the \cite{Somerville99} SAM to explore the accuracy of SED fitting in recovering the physical properties of Lyman-break galaxies (LBGs).
They find that stellar masses are, on average, well recovered for LBGs as they are underestimated, on average, by less than $0.1 \, \mathrm{dex}$. They attribute this success
to two competing factors. Younger stars can mask the presence of older stars in LBG SEDs in some cases, leading to an underestimate of the total stellar mass. However, they 
also find that there is a tendency for SED fitting to overestimate the age of LBGs. This typically leads to overestimating the stellar masses of some galaxies. 

\cite{Pforr12} combine SED fitting with the GalICS SAM \citep{Hatton03}. They find that, in general, stellar masses are 
slightly underestimated when using standard exponentially declining SFHs in the SED fitting process. They demonstrate that this problem can be resolved by adopting exponentially 
increasing SFHs for star-forming galaxies. They conclude that stellar masses are recovered almost perfectly at redshifts $z \in 2, 3$, where the allowed distribution of galaxy ages 
is fairly narrow. At lower redshifts, however, they find that the stellar masses of star-forming galaxies can be underestimated by up to $0.6 \, \mathrm{dex}$ as a result of discrepancies between 
the estimated and true SFHs. They explain that this is caused by the larger range of possible ages at low redshift combined with degeneracies between age and dust. Finally, they also 
show that this problem can be circumvented by choosing to ignore dust reddening when estimating the stellar masses of star-forming galaxies at low redshift. This prevents the SED 
fitting procedure from fitting an unrealistically young age coupled with a large amount of dust reddening.

A key difference in our methodology compared to that of \citeauthor{Pforr12} and \citeauthor{Lee09} is that we use a physically motivated model for attenuation by dust. 
\citeauthor{Pforr12} and \citeauthor{Lee09} instead adopt empirical dust attenuation laws to calculate model galaxy SEDs, corresponding physically to a uniform foreground 
screen of dust placed between the galaxy and an observer. The difference between a foreground dust screen model and the physically motivated radiative transfer 
calculation performed in our analysis turns out to be very significant for our results on stellar mass estimation. It should also be noted that it is not our intent to follow 
\cite{Pforr12} in attempting to quantify the exact level of random and systematic errors on stellar mass estimates for an exhaustive range of possible SED fitting configurations 
and filter sets. This is because any quantitative results derived from the approach of combining SED fitting with theoretical models may be sensitive to the degree to which a 
given model can represent the true galaxy population. Instead, we attempt to provide a detailed explanation of the different error sources we encounter when considering the 
overall galaxy population over a wide range of redshifts. This is achieved by isolating the different factors responsible for biasing stellar mass estimates in a specific set of 
idealized test cases.

The outline of this paper is as follows. In Section~\ref{FITTINGSECTION}, we introduce broad-band SED fitting, explain some of the underlying assumptions that
are involved in the process and outline the parameter choices we make in this study. Section~\ref{GALFORMSECTION} gives a brief overview of \galform
and the specific models which we use in this study. We also explain how intrinsic galaxy properties are transformed into observables in the context of the assumptions made
in SED fitting. In Section~\ref{MassRecoverySection}, we present results of performing SED fitting on model galaxies, focusing on exploring the systematics that affect
the recovery of the stellar masses of individual galaxies. We present results for the stellar mass function over a range of redshifts in Section~\ref{MassFunctionSection}. Finally, we discuss and summarize
our results in Section~\ref{DiscussionSection} and Section~\ref{ConclusionsSection}.

\section{Broad-band SED fitting}
\label{FITTINGSECTION}

We seek to understand the relationship between the stellar mass estimated from observations using SED fitting and the stellar mass predicted by theoretical models.
Instead of using broad-band photometry from observed galaxies, we fit the broad-band magnitudes of model galaxies 
predicted by the semi-analytic model \galform. The precise details of the method used to perform SED fitting in different observational studies typically vary 
very little. Detailed descriptions and discussion of the method can be found in \cite{Bolzonella00}, \cite{Salim07}, \cite{Walcher11} and \cite{Taylor11}. In this section we provide 
an overview of SED fitting as a method of estimating stellar mass. We also describe our parameter choices for the SED fitting procedure used in this study.

\subsection{Overview}

Broad-band SED fitting works by comparing a grid of template galaxy SEDs to observational data. Typically, a maximum-likelihood method is then used to decide which 
template best fits the data \cite[although see][for a discussion of alternative statistical techniques]{Taylor11,Salim07}. This is achieved by first minimizing
$\chi^2$ for each template SED, then choosing the best-fitting template SED with the smallest associated $\chi^2$ value. This corresponds to choosing the mode of the likelihood
distribution. $\chi^2$ is calculated by summing over all available photometric bands using

\begin{equation}
\chi^2 = \sum_{n}\left[\frac{F_{\mathrm{galaxy,}n}-s\,F_{\mathrm{temp,}n}}{\sigma_n}\right]^2,
\label{Eq.chi}
\end{equation}

\noindent where $F_{\mathrm{galaxy,}n}$ and $F_{\mathrm{temp,}n}$ are the fluxes of the galaxy and template, respectively, in the $n$th band, $s$ is a normalization factor and $\sigma_n$ 
is the $1 \sigma$ flux error associated with a galaxy in the $n$th band. The normalization factor $s$ is calculated such that $\chi^2$ is minimized for each template SED using

\begin{equation}
s = \frac{\displaystyle \sum_{m}\left[\frac{ F_{\mathrm{galaxy,}m} \, F_{\mathrm{temp,}m} } {\sigma_{m}^2}\right]}     {\displaystyle  \sum_{m}\left[\frac{F_{\mathrm{temp,}m}}{\sigma_m}\right]^2 },
\label{Eq.s}
\end{equation}

\noindent where a choice can be made regarding which bands are included in the summation. We choose to follow standard practice by simply summing over all available photometric bands, as in
Eq.~\ref{Eq.chi}. 
The stellar mass of each galaxy is then calculated by multiplying the stellar mass of the template by the normalization factor, $s$. This means that the stellar mass is
estimated through normalization over the entirety of the observed galaxy SED, weighted by the error in each band.

Template galaxies SEDs are generated using publicly available SPS models \cite[e.g.][]{Bruzual03, Maraston05, Conroy09}. SPS models predict the spectra of simple 
stellar populations (SSPs), a group of stars with the same age and metallicity and a distribution of initial masses given by the stellar initial mass function (IMF). 
These SSP spectra are then convolved with an assumed 
parametrization for the typical SFH of a galaxy. It is well established that various degeneracies make it very difficult to place strong constraints on galaxy SFHs from 
photometric data alone, unless strong priors are adopted \cite[e.g.][]{Maraston10}. It is therefore standard practice to assume a simple parametrization for the SFH which 
can represent a broad range of possible SFHs without creating a prohibitively large parameter space over which to search.
By far the most common choice of parametrization used for the SFH of low and intermediate redshift galaxies is an exponentially declining SFH. It should be noted, however, that 
there are numerous studies which have advocated alternatives, particularly for high redshift galaxies \cite[e.g.][]{Lee10, Michalowski12, Pacifici13}. The
exponentially declining SFH is parametrized by the time since the onset of star-formation, $t_{\mathrm{age}}$, and the $e$-folding time scale, $\tau$. SPS models also output 
the mass returned from a SSP back into the interstellar medium (ISM) as a function of age, which in turn is used to predict the $M/L$ ratio of each template galaxy. To
reduce the size of the parameter space that needs to be searched over, it is typically assumed that all of the stars in each template galaxy have a single
stellar metallicity, $Z_\star$. Furthermore, the number of metallicity points in the parameter grid is usually very small due to the sparse metallicity grid made
available for publicly available SPS models. For example, there are only 5 metallicities available for the \cite{Bruzual03} (BC03) SPS model. Unlike what is done in theoretical models, it is 
not standard practice to interpolate between metallicities in SED fitting.

\subsection{Dust attenuation}
\label{DustAttenuationSED}

Observed galaxy SEDs are a product of the intrinsic galaxy SED produced by stellar emission which is then attenuated according to radiative transfer through intervening gas and
dust. It is standard practice in SED fitting to account for absorption by neutral hydrogen in the IGM using the prescription from \cite{Madau95}. This is adopted in both
\galform and in the SED fitting procedure used in this study. Attenuation by dust in the ISM of galaxies is a substantially more complex radiative transfer problem. Dust in
galaxies can be concentrated around young stars or distributed diffusely throughout regions of the ISM. Given the lack of information available on the relative spatial distribution
of stars and dust in distant galaxies, attenuation by dust it usually accounted for in SED fitting using the empirical Calzetti dust attenuation law \citep{Calzetti94,Calzetti00}. 
The Calzetti attenuation law has a fixed shape which is different from the dust extinction curve in the local ISM.
The star-dust geometry that is implicitly assumed when applying the Calzetti law corresponds physically to a uniform, foreground dust screen placed between the observer and the stellar 
populations in a given galaxy. Making this assumption has the advantage for SED fitting in that the Calzetti is consequently only characterized by only a single parameter, the reddening 
$E(B-V)$, defined as the difference between the observed and intrinsic $B-V$ colour. The shape of the Calzetti law was derived from a sample of 30 local starbursts \citep{Calzetti94} 
and the normalization was derived from a sub-sample of only 4 local starbursts \citep{Calzetti00}.

A number of studies have attempted to assess how well the Calzetti law can reproduce the attenuation curves of different galaxy types. At low redshift, \cite{Wild11} apply a pair-matching 
technique to study the shape of the attenuation curves of SDSS spirals. They find evidence that the shape of the optical dust attenuation curves of local star-forming galaxies is strongly 
dependent on galaxy inclination. Specifically, they show that face-on spirals have steeper optical attenuation curves than the Calzetti law, whereas edge-on spirals are consistent with the 
Calzetti law. They also find that the slopes of the near-infrared (NIR) attenuation curves are consistent with Milky-Way extinction or 
Calzetti law attenuation curves, independent of inclination. In the UV, they find evidence for a bump in the attenuation curve of spirals at $2175 \, \mathrm{\AA}$ \cite[see also][]{Conroy10dust}. 
This feature is absent from the Calzetti law and could have a significant impact on the interpretation of the properties of high redshift galaxies \citep{Gonzalez-Perez13}.

Another method that is used to investigate the dust attenuation properties of galaxies involves measuring how the ratio of far-infrared (FIR) to ultraviolet (UV) flux, IRX, varies as a function of the 
UV spectral slope, $\beta$ \cite[e.g.]{Bell02, Goldader02, Howell10, Murphy11, Penner12}. The position of galaxies on the IRX-$\beta$ plane can then be compared with the relationship derived 
for local starbursts \citep{Meurer99, Overzier11}. The \cite{Meurer99} relation was derived from the same galaxy sample used to derive the Calzetti law and so this comparison tests whether
the Calzetti law is applicable to objects other than modestly starbursting local galaxies. \cite{Bell02} was the first to show that local star-forming galaxies lie below this relation such that 
there is less UV attenuation for a given value of $\beta$. \cite{Goldader02} were the first to show that local ultraluminous infrared galaxies lie above this relation such that the Calzetti law underestimates 
the total UV attenuation for these objects.

Various studies have also applied this method to high redshift star-forming galaxy samples \cite[e.g.][]{Murphy11, Buat12, Reddy12, Penner12}. This exercise is difficult because FIR SEDs are typically 
available only for the most extreme dusty galaxies at higher redshifts, although a stacking analysis can ameliorate this problem \citep{Reddy12}. There is evidence for consistency between the 
high redshift and local IRX-$\beta$ relations \cite[e.g.][]{Reddy10, Reddy12}, although it has also been argued that specific object classes can be offset from the local relation 
\cite[e.g.][]{Murphy11, Penner12}. \cite{Buat12} apply a different approach and analyse a sample of 751 UV-selected galaxies at $z \in 1,2$, fitting the full UV-FIR SEDs using a SED fitting 
procedure that features a generalized form of the Calzetti law \citep{Noll09}. They find evidence for a steeper attenuation curve in the UV than the canonical Calzetti law for $20\%$ of their sample,
and also a UV bump.

\subsection{Filter and parameter choices}

The final step in producing template SEDs involves convolving with the broad-band filters used in given observational data set. Deep multi-wavelength surveys 
such as GOODS have many photometric bands available, spanning all the way from the UV through to the radio. Wide-area surveys on the other hand such as SDSS typically have only optical 
broad-band photometry available. For simplicity, we use a single filter set across a wide range of redshifts with the exception of Section~\ref{LBGsection} where we consider LBG samples. 
Filters blueward of the Lyman limit at $912 \mathrm{\AA}$ in the rest frame are excluded from the fitting process. We do not include artificial redshift or flux errors, setting
$\sigma_n$ in Eq.~\ref{Eq.chi} to $10 \%$ of the model galaxy flux $F_{\mathrm{galaxy,}n}$ for each band. These error sources become important at high redshift but can be understood 
without the need for a theoretical model and are not particularly relevant for understanding errors in stellar mass estimates associated with the assumptions made in SED fitting. However, 
ignoring them entirely means that any quantification of the errors in stellar mass estimates given in this paper should be considered as lower limits. We do consider the effect
of artificially perturbing model fluxes when exploring LBG samples in Section~\ref{LBGsection}, where it becomes important to include detection criteria in order to robustly compare model predictions 
with observational data.

In this study we use two filter and SED fitting parameter sets, as outlined in Table~\ref{SEDfittingParameters}. A common feature of both sets is that we use \cite{Bruzual03} SPS models and the Calzetti
law for SED fitting. $t_{\mathrm{age}}$
is always constrained to be less than the age of the Universe at the given redshift.
We refer to the first parameter set as the standard parameter grid because it is designed to be broadly representative of the choices made in observational studies of low to intermediate redshift galaxies
where Spitzer IRAC imaging between $3.6 \mathrm{\mu m}$ and $8\mathrm{\mu m}$ is often available \cite[e.g.][]{Ilbert10, Santini12, Mortlock11}.
It uses the exponentially declining SFH typically used for galaxies at low and intermediate redshift and is therefore characterized by $t_{\mathrm{age}}$, 
$\tau$, $Z_\star$ and $E(B-V)$. We use a Salpeter IMF for this parameter set despite the fact that the \galform models we consider typically use a Kennicutt IMF to demonstrate how the 
systematic uncertainty on the IMF compares against other sources of error in stellar mass estimation. We modify this choice of IMF in the templates to a Chabrier IMF in Section~\ref{MassFunctionSection} 
where we consider model predictions of the stellar mass function. This choice is made so that the stellar masses of model galaxies estimated from SED fitting are consistent with the observational studies 
with which we compare.
 
We also use a second parameter set, deliberately constructed to closely resemble the choices made by \cite{Lee12}, who use SED fitting to estimate the stellar masses of LBGs at $z=4$ and $z=5$. 
We refer to this parameter set as the LBG parameter grid and use it in Section~\ref{LBGsection}. It uses an 
exponentially declining SFH, including the limit of $\tau \to \infty$, corresponding to a constant SFH. When considering LBGs, it is important to consider both 
the effect of photometric errors and non-detections where galaxies drop below the sensitivity limit of the survey. Instead of setting $\sigma_n$ in Eq.~\ref{Eq.chi} to $10 \%$ of 
the model galaxy flux $F_{\mathrm{galaxy,}n}$, in this case we use the 5-sigma limiting magnitudes listed in Table 1 of \cite{Lee12} to determine $\sigma_n$. We also consider the effect of artificially perturbing fluxes using
a Gaussian distribution with $\sigma$ again taken from Table 1 in \cite{Lee12}. To facilitate a self-consistent comparison, we follow the same procedure for non-detections described in \cite{Lee12}, where non-detected bands 
are used as upper limits in Eq.~\ref{Eq.chi} if $s \, F_{\mathrm{temp,}n}$ exceeds the upper limit in that band. We use the LBG dropout selection criteria (both colour and $S/N$ selection) given 
by Equations 1-13 in \cite{Stark09}. These criteria select $B_{435}$, $V_{606}$ and $i_{775}$ dropouts to create samples of LBGs at $z=4, 5$ and $6$ respectively. 
We use a Chabrier IMF for this parameter set. The allowed parameter values for both parameter sets are listed in Table~\ref{SEDfittingParameters}.

\begin{table}
\begin{tabular}{|c|l|}
\hline
\multicolumn{2}{c}{Standard Parameter Grid} \\
\hline
Filters & $B_{435}$, $V_{606}$, $R$, $i_{775}$, $z_{850}$, $J$, $H$, $K$, $3.6$, $4.5$, $5.8$, $8.0 \mathrm{\mu m}$ \\
IMF & Salpeter \\
$t_{\mathrm{age}}/\mathrm{Gyr}$ & 0.1, 0.11, 0.13, 0.14, 0.16, 0.18, 0.2, 0.23, ...  \\
$\tau / \mathrm{Gyr}$ & 0.1, 0.3 0.6, 1, 2, 3, 4, 5 7, 9, 13, 15, 30 \\
$Z_\star / \mathrm{Z_{\odot}}$ & 2.5, 1, 0.4, 0.2, 0.02 \\
$E(B-V)$ & 0, 0.03, 0.06, 0.1, 0.15, 0.2, 0.25, 0.3, ... 1 \\
\hline
\multicolumn{2}{c}{Lyman-Break Galaxy Parameter Grid} \\
\hline
Filters & $B_{435}$, $V_{606}$, $i_{775}$, $z_{850}$, $J$, $H$, $K$, $3.6$, $4.5$, $5.8 \mathrm{\mu m}$ \\
IMF & Chabrier \\
$t_{\mathrm{age}}/\mathrm{Gyr}$ & 0.1, 0.11, 0.13, 0.14, 0.16, 0.18, 0.2, 0.23, ...  \\
$\tau / \mathrm{Gyr}$ & 0.1, 0.2 0.3, 0.4, 0.6, 0.8, 1.0 , $\infty$\\
$Z_\star / \mathrm{Z_{\odot}}$ &  1, 0.2 \\
$E(B-V)$ & 0, 0.025, 0.05, 0.075 ... 0.95 \\
\end{tabular}
\caption{Parameter grids for SED fitting. The top section outlines the standard parameter grid we use for the majority of our results. The bottom section
outlines the parameter grid used in Section~\ref{LBGsection} for exploring LBG sample selection. $t_{\mathrm{age}}$
is the time since the onset of star-formation and $\tau$ is the $e$-folding time-scale for an exponentially decreasing SFH.
$Z_\star$ is the stellar metallicity and $E(B-V)$ is the colour excess which characterizes the Calzetti dust attenuation law.}
\label{SEDfittingParameters}
\end{table}

\section{Modelling Hierarchical Galaxy Formation}
\label{GALFORMSECTION}

In this section we provide a brief description of the aspects of the \galform semi-analytic model
which are relevant to this work. An introduction to the model and the associated underlying physics 
can be found in \cite{Cole00}, \cite{Baugh06} and \cite{Benson10}. Briefly, \galform connects the 
properties of galaxies to a given cosmological model by coupling dark matter halo merger trees to a 
set of continuity equations that describe the exchange of baryons accreted on to dark matter haloes 
between stellar, cold disk gas and hot halo gas components. The physical processes that determine
the form of these continuity equations include shock heating and subsequent radiative cooling of
accreted gas onto galaxy disks, quiescent star-formation in galaxy disks, chemical enrichment of
the ISM, the ejection of cold gas and metals by supernova feedback, the suppression of gas cooling 
by AGN and photoionization feedback and disk instabilities and galaxy mergers that can trigger both 
spheroid formation and bursts of star-formation.
It is important to note that various versions of \galform have appeared in the 
literature which we refer to as separate models. These models are distinct from each other in that 
they all use different choices for model parameters and in some cases actually include physical 
processes which do not appear in other models.

\subsection{The Lagos12 and Lacey13 models}
\label{ModelDescription}

We adopt the recently developed model described in \cite{Lagos12} (hereafter Lagos12) as the fiducial model to explore in 
this study. The Lagos12 model is the most recent version of the model described in \cite{Lagos11a}, which in turn is a development 
of the model originally described in \cite{Bower06} (hereafter Bower06). The Bower06 model was the first variant of \galform to include the 
effects of AGN feedback shutting down gas cooling in massive haloes. The Lagos12 model is distinct from the Bower06 model in that it includes 
an alternative star formation law for galaxy disks based on an empirical relationship connecting the star formation rate in a galaxy to the molecular-phase 
gas density. The molecular gas fraction is, in turn, related to the mid-plane pressure in the galaxy disk \citep{Blitz06}. This new law is observationally 
motivated and is characterized by parameters which are constrained by observations, greatly reducing the available parameter space in the model.
Other changes with respect to Bower06 include longer time scales for both the minimum and total duration of starbursts 
and different reionization parameters. These changes were made to reconcile the model predictions with observations of 
LBGs \citep{Lacey11}. The model uses SPS files from a 1999 private release 
of the Bruzual \& Charlot model family (BC99) and assumes a Kennicutt IMF \citep{Kennicutt83}. 
The BC99 SPS models represent an intermediate step between the public model releases from \cite{Bruzual93} and \cite{Bruzual03} and are found to be very similar
to the BC03 SPS models.
Compared to the $x=1.35$ Salpeter IMF used in the SED fitting procedure, the Kennicutt IMF used in the Lagos12 model has the same mass range ($m_\star \in 0.1,100 \, \mathrm{M_\odot}$) but has a 
steeper slope of $x=1.5$ and a break in the power law at $m_\star = 1 \mathrm{M_\odot}$, below which the slope is $x=0.4$. 
The IMF slope $x$ is defined by $\frac{\mathrm{d}N(m)}{\mathrm{d}\ln{m}} \equiv m^{-x}$.
The lack of a power law break at $0.1 \, \mathrm{M_\odot}$ means that the Salpeter IMF has an overabundance of low-mass stars compared to the Kennicutt IMF, 
resulting in higher $M/L$ ratios.

To help explore if certain aspects of our results are model dependent, we also consider the model presented in 
Lacey et al. (in preparation) (hereafter Lacey13). The Lacey13 model is a hybrid of the Bower06 model family and the model 
from \cite{Baugh05}. It includes AGN feedback, starburst events triggered by disk instabilities and galaxy mergers,
the star formation law described in \cite{Lagos11a} and a non-universal IMF. We choose this model as a comparison
to the Lagos12 model because of several differences between the models which are relevant for SED fitting. The non-universal 
IMF used in the Lacey13 model consists of a Kennicutt IMF for star formation in disks and a top-heavy IMF with slope $x=1$ in starbursts.
It should be noted that the slope of the top-heavy IMF used in the Lacey13 
model is less extreme than the $x=0$ top-heavy IMF required in \citeauthor{Baugh05} to match the number counts of submillimetre galaxies. 
The Lacey13 model generates SEDs using the \cite{Maraston05} (hereafter MA05) SPS model. There has been a considerable amount of debate in the
literature as to whether the luminosity of thermally-pulsating asymptotic giant (TP-AGB) stars featured in the MA05 model is accurate
\cite[e.g.][]{Kriek10, Zibetti13}. This has potentially important consequences for the stellar mass inferred
from SED fitting, potentially changing the $M/L$ at NIR wavelengths by as much as $50\%$ for a stellar population
of age $\approx 1 \mathrm{Gyr}$ \citep{Maraston06, Michalowski12}. Although we do not explore this issue in any detail, the debate
surrounding the contribution from TP-AGB stars makes the Lacey13 model a useful comparison to our fiducial model.

\subsection{Calculating intrinsic galaxy SEDs}

The SED fitting procedure described in Section~\ref{FITTINGSECTION} relies on the accuracy of SPS modelling to provide realistic SEDs for simple stellar
populations. The same is true for \galform which uses SPS modelling to predict model galaxy SEDs from the star formation and chemical
enrichment history of each galaxy, as calculated by the model. Compared to SED fitting, which has to assume a parametric form for galaxy 
SFHs and a single metallicity for all of the stars in a given galaxy, \galform self-consistently generates complex assembly
histories for galaxies which include chemical evolution \cite[see][for examples]{Baugh06}. 
There are only a small number of metallicities available for publicly released SPS models.
Therefore, in order to actually use the chemical enrichment history of each galaxy in \galform, the model performs 
linear interpolation in $\mathrm{log}(Z_{\star})$ between the tabulated SSPs. This approach is not applied in standard SED fitting procedures which instead use a 
discrete metallicity grid. Another difference between \galform and SED fitting is that galaxies in \galform are divided 
into disk and bulge components. The net SED of each model galaxy predicted by \galform is therefore the sum of two composite stellar 
populations, each with its separate star formation and chemical enrichment history. Finally, it should also be noted that \galform uses a different
choice with respect to SED fitting regarding the treatment of the recycling of mass and metals. SPS models typically provide estimates of the amount of mass
a SSP recycles back to the ISM as a function of age. This information is used in SED fitting to calculate the best-fitting stellar mass. For reasons of numerical
efficiency, theoretical galaxy formation models, including \galform, typically do not use this information and instead apply the instantaneous recycling 
approximation where mass and metals are instantly returned to the ISM. The amount of mass and metals returned per unit mass of stars formed are both fixed parameters in \galform and are therefore independent of the
age of a galaxy. The exact recycled fractions and yields are calculated, for a given IMF, from the output of a SSP with solar metallicity at an age of 
$10 \, \mathrm{Gyr}$. This has a direct impact on how stellar mass is calculated in \galform and it is expected that this will lead to a 
small, redshift-dependent disagreement with the non-instantaneous recycling scheme employed in SED fitting.

\subsection{Dust attenuation}

The SED fitting procedure and \galform both use the same \cite{Madau95} prescription for absorption and scattering of UV photons
by neutral hydrogen in the IGM. However, there are important differences in the way attenuation by dust is treated.
The SED fitting procedure uses the Calzetti law which includes the assumption of a star-dust geometry corresponding
to a uniform, foreground dust screen, as discussed in Section~\ref{DustAttenuationSED}. \galform performs
a more physical calculation of radiative transfer for a realistic geometry of the stars and dust. It models dust as a two 
phase medium separated into diffuse dust in the 
ISM and compact dust clouds that enshroud star-forming regions \citep{Silva98}. For a detailed description of this dust attenuation model, 
see \cite{Lacey11} and references therein and also Lacey, Baugh \& Frenk (in preparation). We provide a qualitative 
overview of the model here, focusing on aspects of the modelling that are particularly relevant to our analysis. 
We use the standard terminology whereby extinction curves describe the absorption and scattering out of the sightline 
to a single star and attenuation curves describe the total absorption and scattering both into and out of all sightlines to 
an extended stellar distribution on the sky. We characterize attenuation curves with the effective optical depth, 
$\tau_{\mathrm{eff}}$, as a function of wavelength $\lambda$. The effective optical depth, $\tau_{\mathrm{eff,}\lambda}$, is defined by

\begin{equation}
\tau_{\mathrm{eff,}\lambda} = -\ln(F_{\mathrm{atten,}\lambda}/F_{\mathrm{intrin,}\lambda}),
\label{Eq.effoptdepth}
\end{equation}

\noindent where $F_{\mathrm{atten,}\lambda}$ and $F_{\mathrm{intrin,}\lambda}$ are, respectively, the attenuated and intrinsic galaxy fluxes at a given wavelength and
$F_{\mathrm{atten,}\lambda}$ is calculated from a radiative transfer model.

The starting point for the dust model used in \galform is to calculate the total dust mass in each galaxy. This is
calculated by assuming that the ratio of mass in dust to metals in the cold gas is a constant and that this ratio follows the value 
inferred for the local ISM \citep{Savage79}. Dust is then divided into diffuse and compact birth cloud components. The relative
fraction of dust mass in each component is a model parameter. The fraction in diffuse dust, $f_{\mathrm{diffuse}}$, is set to $0.75$ in the
Lagos12 model and $0.5$ in the Lacey13 model. Both dust components use an input Milky-Way extinction curve as a starting
point to calculate the resultant attenuation curves of each galaxy by radiative transfer. It is important to realise that inclination and geometric
effects can lead to total attenuation curves which are very different from the input extinction curve \cite[e.g.][]{Gonzalez-Perez13}.

The spatial distribution of diffuse dust is assumed to follow an exponential disk profile that traces the stellar disk, both in
radial and vertical scale-length.
The effective optical depth associated with diffuse dust is calculated by interpolating between the tabulated radiative transfer 
calculations performed by \cite{Ferrara99}. \cite{Ferrara99} calculate the effective optical depth of disk-bulge systems as a 
function of wavelength, galaxy inclination, face-on extinction optical depth in the $V$-band, $\tau_{V0}$, and disk-to-bulge scale-length ratio.
$\tau_{V0}$ is calculated directly from the density of dust at the centre of the disk, using the local ISM dust to metals ratio, and so scales 
with the surface density of diffuse dust in the galaxy disk as

\begin{equation}
\tau_{V0} \propto f_{\mathrm{diffuse}} M_{\mathrm{cold}} Z_{\mathrm{cold}} / {r_{\mathrm{disk}}}^2,
\label{Eq.tauv0}
\end{equation}

\noindent where $M_{\mathrm{cold}}$ and $Z_{\mathrm{cold}}$ are the mass and metallicity of cold gas in the galaxy disk and $r_{\mathrm{disk}}$ is the radius of the
galaxy disk. It should be noted that no allowance is made in the modelling of diffuse dust for any differences in the relative spatial 
distributions of young and old stars in the galaxy disk, although this is accounted for with the second, compact dust cloud component. It is normally 
assumed that there is no dust in galaxy bulges. An exception is made, however, for diffuse dust associated with gas forming stars in starbursts 
triggered by mergers or disk instabilities. In these starbursting systems, the attenuation by diffuse dust is approximated by temporarily treating the bulge 
as a disk when using the results from the \cite{Ferrara99} radiative transfer calculations. We discuss some of the advantages and potential problems associated with the
way that diffuse dust is modelled in \galform, in the context of our results, in Section~\ref{DiscussionSection}.

The second dust component in \galform represents dust in dense molecular clouds enshrouding star-forming regions. As such, it generally affects the light
emitted only by young stars which in turn are assumed to escape the dense dust clouds over a fixed time scale, which is a model parameter. The cloud component 
is therefore more significant in actively star-forming galaxies and starbursts where very young stellar populations can dominate large parts of the overall 
galaxy SED. The escape time is set to $1 \, \mathrm{Myr}$ for the Lagos12 model and $1 \, \mathrm{Myr}$ for the Lacey13 model.
The clouds are modelled as being spherically symmetric with uniform density and a mass of $10^6 \mathrm{M_\odot}$ and a radius of $16 \mathrm{pc}$. The 
enshrouded stars are placed at the centre. Attenuation from this simple geometry can be evaluated analytically. For a more detailed description of this
aspect of the calculation, see Lacey, Baugh \& Frenk (in preparation).  

The resultant combination of the diffuse and compact dust components attenuating the overall galaxy SED increases the level of physical realism beyond
what is represented by the Calzetti law used in SED fitting. We pay particular attention to this in Section~\ref{dustresults} but it should be noted
that a full exploration of how the dust modelling used in \galform compares to the empirical relations used in observational studies is beyond the scope of
this paper. For more information, see \cite{Gonzalez-Perez13} for a discussion of how model predictions derived using this dust modelling approach compare 
with observations of LBGs.

\section{Stellar Mass Recovery}
\label{MassRecoverySection}

In this section we first examine how accurately SED fitting can recover the stellar masses of a volume limited sample of model galaxies 
predicted by the Lagos12 model at a selection of redshifts. We then attempt to isolate and explain the various different effects which
affect the accuracy of stellar mass estimation. In this section, we use the standard SED fitting parameter grid and filter set described in the 
top half of Table~\ref{SEDfittingParameters}. 
The number of model galaxies considered at each redshift is of the order of $10^5$, such that the galaxy population is well represented.

\subsection{Overview}

\begin{figure*}
\begin{center}
\includegraphics[width=40pc]{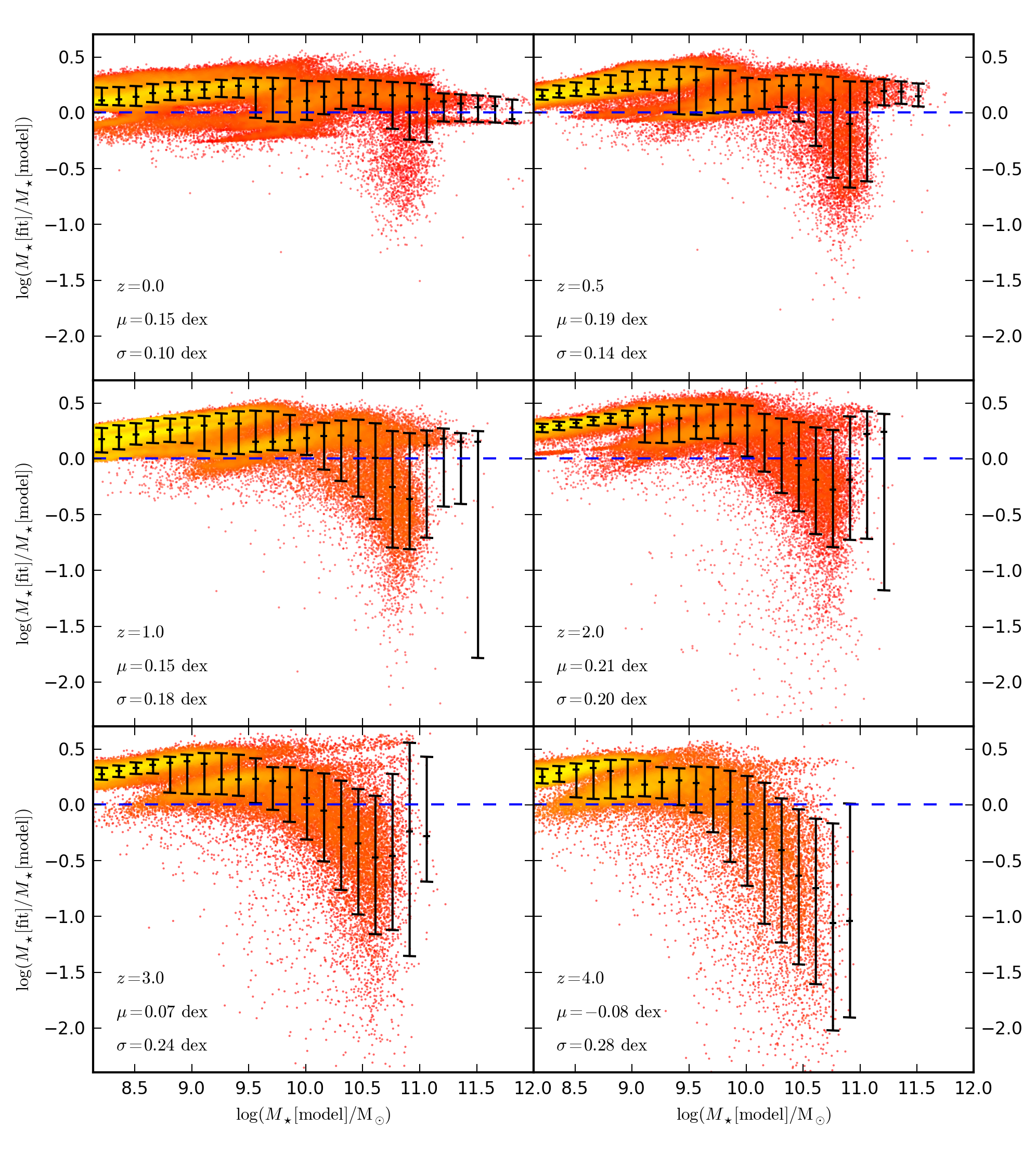}
\caption{The log of the ratio of the stellar mass estimated using SED fitting to the true stellar mass in the Lagos12 model, plotted as a function of the true stellar mass. 
Each panel corresponds to a different redshift as labelled. 
The coloured points represent individual model galaxies. 
The point colours are scaled logarithmically with the local point density in the panel, from red at low density to yellow at high density.
The black points and corresponding error bars show the median, $10$ and $90$ percentiles of the distribution in bins of true stellar mass.
$\mu$ is the mean median offset and $\sigma$ is half the mean $68\%$ range of the distribution. 
For reference, the blue dashed line shows the locus of equality between estimated and true stellar mass.}
\label{mass.recovery.evo.Lagos12}
\end{center}
\end{figure*}

Fig.~\ref{mass.recovery.evo.Lagos12} shows the ratio of the stellar mass estimated using SED fitting, $M_\star \mathrm{[fit]}$ to the true stellar mass in the Lagos12 model,
$M_\star \mathrm{[model]}$, plotted as a function of $M_\star \mathrm{[model]}$ for a selection of redshifts. We choose to show individual galaxies colour coded by the density of points at a 
given position on the plane. We also show the 10, 50 and 90 percentile ranges of the distribution. This approach shows the broad trends in the overall distribution whilst
still highlighting the presence of any unusual features or outliers. We quantify the distributions in each panel using two simple statistics in order to facilitate 
a rough quantitative comparison with other results presented in this section. We define $\mu$ as the mean value of the median offset in $\log_{10}(M_\star \mathrm{[fit]} / M_\star \mathrm{[model]})$
calculated for each bin in $M_\star \mathrm{[model]}$. We define $\sigma$ as half of the mean value of the 68\% range in $\log_{10}(M_\star \mathrm{[fit]} / M_\star \mathrm{[model]})$
calculated for each bin in $M_\star \mathrm{[model]}$. If there is no dependence of the scatter and median offset on $M_\star \mathrm{[model]}$, then $\mu$ and $\sigma$ quantify exactly the average systematic 
and random errors which affect the stellar mass estimation. 

At face value, the results shown in Fig.~\ref{mass.recovery.evo.Lagos12} indicate that the accuracy of the stellar masses estimated using SED fitting is very poor, 
particularly at high redshift. It should be noted, however, that we have deliberately chosen to assume a Salpeter IMF in our SED fitting procedure despite the fact that the Lagos12 
model uses a Kennicutt IMF. The difference in $M/L$ ratio between the Salpeter and Kennicutt IMFs can account for the systematic offset in $M_\star \mathrm{[fit]} / M_\star \mathrm{[model]}$ 
seen for low mass galaxies. However, the IMF mismatch cannot explain the behaviour displayed for massive galaxies, particularly at high redshift. These galaxies display a huge scatter in 
$M_\star \mathrm{[fit]} / M_\star \mathrm{[model]}$. Specifically, there seems to be a population of massive galaxies where the stellar mass is significantly underestimated. The 
medians and percentiles of the overall distribution show that this is an outlying population at low redshift. However, at high redshift, it is apparent that the stellar masses of almost 
all of the most massive model galaxies is significantly underestimated. In the most extreme individual cases, the stellar mass can be underestimated by factors greater than a hundred. 
Finally, the distributions display a level of bimodal behaviour 
which can be seen by eye from the point density distribution indicated by the colour scheme. This is easier to see in the higher redshift panels. The two peaks of the bimodal feature are 
typically offset in $\log_{10}(M_\star \mathrm{[fit]} / M_\star \mathrm{[model]})$ by $\approx 0.25 \, \mathrm{dex}$. This is significant and clearly undesirable.

\begin{figure*}
\begin{center}
\includegraphics[width=40pc]{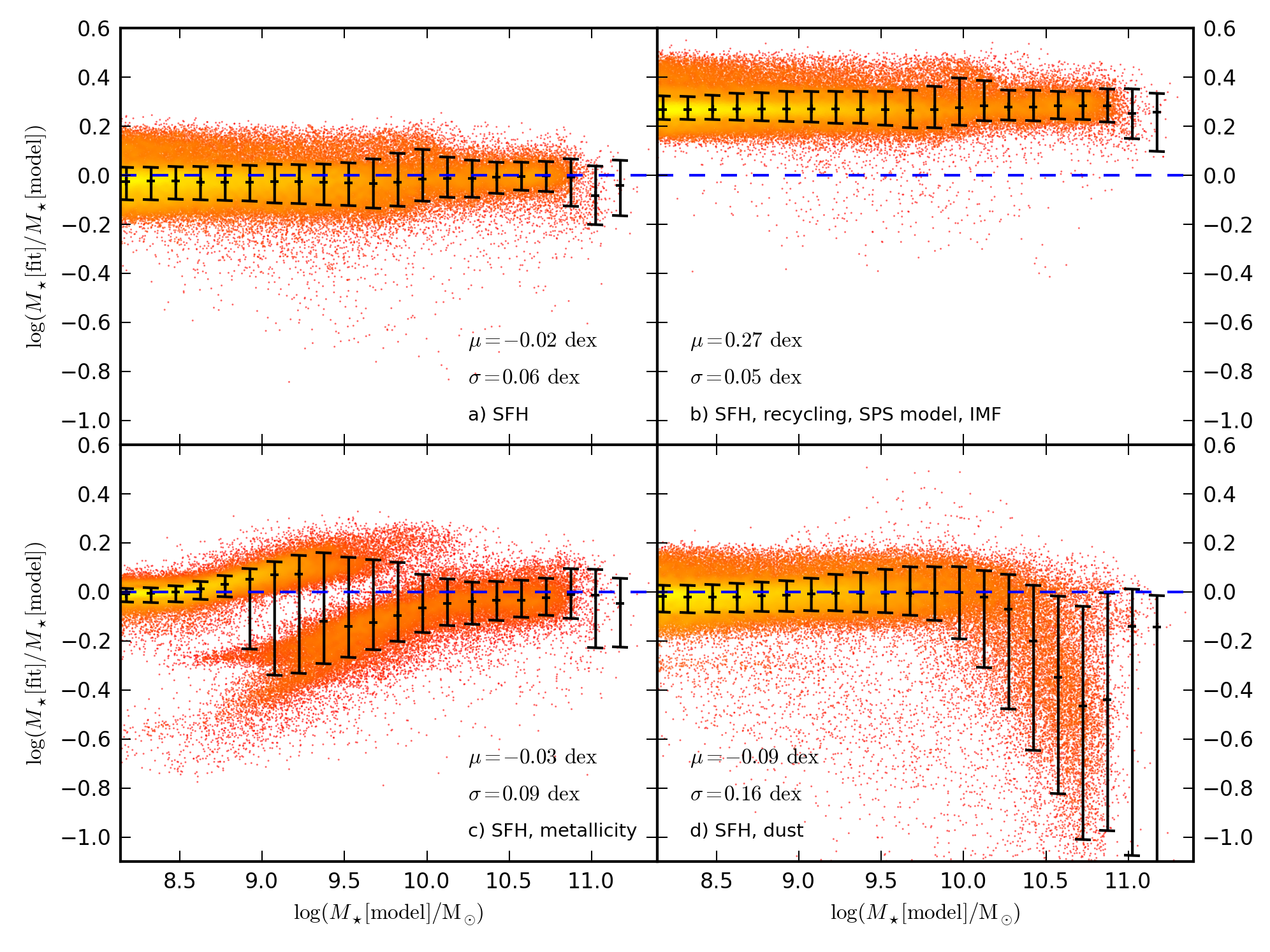}
\caption{The log of the ratio in the stellar mass estimated using SED fitting to the true stellar mass in the Lagos12 model at $z=2$, plotted as a function of the true stellar mass. 
Formatting of points and symbols is the same as in Fig.~\protect{\ref{mass.recovery.evo.Lagos12}}.
The different panels show the distribution for different variations of both SED fitting and \galform.
{\it a)} The SFHs of galaxies are the only factor which can vary in the SED fitting process. 
No dust extinction is applied to model galaxy SEDs in \galform and $E(B-V)=0$ is applied as a constraint in the SED fitting procedure. 
$Z_\star = Z_\odot$ is applied as a constraint in both \galform and the SED fitting.
The SPS model used by \galform is changed to BC03 with a Salpeter IMF (in order to be consistent with the SED fitting) and the instantaneous recycling approximation is used in the SED fitting procedure (to be consistent with \galform).
{\it b)} SFHs, recycling, SPS models and the IMF are the only factors in the SED fitting process. Dust and metallicity related effects are removed as in Panel {\it a)}.
{\it c)} SFHs and metallicity are the only factors in the SED fitting process. Dust, recycling, SPS models and IMF related effects are removed as in Panel {\it a)}.
{\it d)} SFHs and dust are the only factors in the SED fitting process. Metallicity, recycling, SPS model and IMF related effects are removed as in Panel {\it a)}.}
\label{mass.recovery.flythrough.Lagos12}
\end{center}
\end{figure*}

There are a number of different factors in the SED fitting calculation that could combine to produce the behaviour shown in Fig.~\ref{mass.recovery.evo.Lagos12}. 
Therefore, it is useful to modify the SED fitting procedure and \galform in order to isolate how each factor of the calculation contributes to this overall behaviour. 
In Fig.~\ref{mass.recovery.flythrough.Lagos12}, we show how the distribution of estimated over true stellar mass changes with the inclusion or exclusion of these factors for the 
Lagos12 model. We adopt a fiducial redshift of $z=2$ for this exercise.
Fig.~\ref{mass.recovery.flythrough.Lagos12}a shows the case where \galform and the SED fitting procedure have been stripped down to the
point where effectively only the SFH is being fit for each model galaxy. This is achieved by removing all of the effects associated with dust attenuation, chemical
enrichment, recycling and the choice of SPS model and IMF. SPS model, recycling and IMF related effects are removed simply by making the two calculations consistent. Specifically, the 
SPS model used by \galform is changed to BC03 with a Salpeter IMF and the instantaneous recycling approximation is adopted in the SED fitting procedure. We remove chemical 
enrichment effects by forcing SED calculations in both \galform and the SED fitting procedure to use solar metallicity. Dust effects are removed by setting $E(B-V)=0$ as a 
constraint in the SED fitting procedure and by using the unattenuated fluxes for model galaxies from \galform. From this simplified case, the other panels show how the distribution
changes with the reintroduction of the various aspects of the calculation that were removed in Fig.~\ref{mass.recovery.flythrough.Lagos12}a. Each aspect is reintroduced in isolation.

The remainder of this section is outlined as follows. In Section~\ref{SPSsection}, we discuss the role of SFHs, SPS models, recycling and the choice of IMF on the inferred stellar mass. 
In Section~\ref{Wavelength}, we explore how our results are affected by wavelength coverage. Section~\ref{MetallicitySection} and Section~\ref{dustresults} discuss the impact of metallicity and dust respectively.
In Section~\ref{AlternativeSection}, we extend our analysis to the Lacey13 model to explore the model dependence of our results.

\subsection{SFHs, recycling, SPS models and the IMF }
\label{SPSsection}

\begin{table}
\begin{tabular}{|c|l|l|l|l|l|l|}

z & 0 & 0.5 & 1 & 2 & 3 & 4 \\
\hline
\multicolumn{7}{c}{\galform: BC03 SPS, Salpeter / SED fitting: BC03 SPS, Salpeter, IRA} \\
$\mu /\mathrm{dex}$ & -0.01 & -0.01 & -0.03 & -0.02 & -0.01 & -0.01 \\
$\sigma /\mathrm{dex}$ & 0.01 & 0.03 & 0.04 & 0.06 & 0.05 & 0.05 \\
\hline
\multicolumn{7}{c}{\galform: BC03 SPS, Salpeter / SED fitting: BC03 SPS, Salpeter, NIRA} \\
$\mu /\mathrm{dex}$ & -0.01 & -0.01 & -0.01 & 0.01 & 0.04 & 0.05 \\
$\sigma /\mathrm{dex}$ & 0.01 & 0.03 & 0.04 & 0.05 & 0.05 & 0.05 \\
\hline
\multicolumn{7}{c}{\galform: BC99 SPS, Salpeter / SED fitting: BC03 SPS, Salpeter, NIRA} \\
$\mu /\mathrm{dex}$ & -0.03 & -0.01 & -0.02 & 0.00 & 0.03 & 0.06 \\
$\sigma /\mathrm{dex}$ & 0.04 & 0.04 & 0.05 & 0.05 & 0.05 & 0.05 \\
\hline
\multicolumn{7}{c}{\galform: BC99 SPS, Kennicutt / SED fitting: BC03 SPS, Salpeter, NIRA} \\
$\mu /\mathrm{dex}$ & 0.26 & 0.26 & 0.26 & 0.27 & 0.30 & 0.31 \\
$\sigma /\mathrm{dex}$ & 0.03 & 0.05 & 0.05 & 0.05 & 0.05 & 0.05 \\

\end{tabular}
\caption{The mean median offset $\mu$ and half the mean $68\%$ range, $\sigma$, of distributions in $\log(M_\star\mathrm{[fit]}/M_\star\mathrm{[model]})$ against $M_\star\mathrm{[model]}$. 
All values listed are for the Lagos12 model in the case where dust effects are ignored, both in the model and in the SED fitting procedure. 
Metallicity effects are also removed by forcing both the model and the SED fitting procedure to use $Z_\star = \mathrm{Z_\odot}$. 
Each column corresponds to a different redshift. Each pair of rows corresponds to a different combination of choices made regarding SPS modelling, the IMF and recycling in \galform and the SED fitting procedure. 
The top pair of rows corresponds to the simplified case where \galform and the SED fitting procedure both use BC03 SPS models, instantaneous recycling (IRA) and a Salpeter IMF.
The second pair of rows corresponds to the case where the SED fitting procedure is changed back to using default non-instantaneous recycling (NIRA).
The third pair of rows corresponds to the case where the Lagos12 model is changed back to using default BC99 SPS models and the SED fitting procedure uses NIRA.
The final pair of rows corresponds to the default case where \galform uses BC99 SPS models, IRA and a Kennicutt IMF.
The corresponding default SED fitting procedure uses BC03 SPS models, a Salpeter IMF and NIRA.}
\label{SPSimfRec}
\end{table}

As discussed earlier, the case presented in Fig.~\ref{mass.recovery.flythrough.Lagos12}a is simplified to the extent where the only difference between SED calculations performed
by \galform and the fitting procedure is in the form of the galaxy SFHs. The SED fitting procedure assumes an exponentially declining SFH characterized by the time since the onset 
of star-formation, $t_{\mathrm{age}}$, and the $e$-folding time scale, $\tau$, whereas \galform self-consistently calculates the SFH of each model galaxy. None of the concerning features 
and trends seen in Fig.~\ref{mass.recovery.evo.Lagos12} are present in Fig.~\ref{mass.recovery.flythrough.Lagos12}a, which instead shows a smooth distribution with a small scatter 
almost centered around the locus of equality between estimated and true stellar mass. The distribution can be completely characterized by the mean offset $\mu = -0.02\, \mathrm{dex}$ 
and the mean spread $\sigma = 0.06\, \mathrm{dex}$ in this idealized case. It is perhaps surprising that, on average, the SED fitting works so well given the diversity of SFHs which
can be predicted in \galform, and it is interesting then to see whether this result is reproduced at other redshifts. We list the $\mu$ and $\sigma$ values for this simplified case for 
other redshifts in the top pair of rows in Table~\ref{SPSimfRec}. It is interesting to see that, averaged over the entire galaxy population, the assumption of an exponentially declining 
SFH has almost no impact on the accuracy of the stellar mass estimation at $z=0$ ($\mu = -0.01 \, \mathrm{dex}$ and $\sigma = 0.01 \, \mathrm{dex}$). In addition, the small scatter seen in 
Fig.~\ref{mass.recovery.flythrough.Lagos12}a at $z=2$ does not increase for higher redshifts. Comparing this level of scatter with that seen in Fig.~\ref{mass.recovery.evo.Lagos12} 
implies that, for our analysis, the assumption of an exponentially declining SFH has a negligible impact on stellar mass estimation. 

\begin{figure}
\includegraphics[width=20pc]{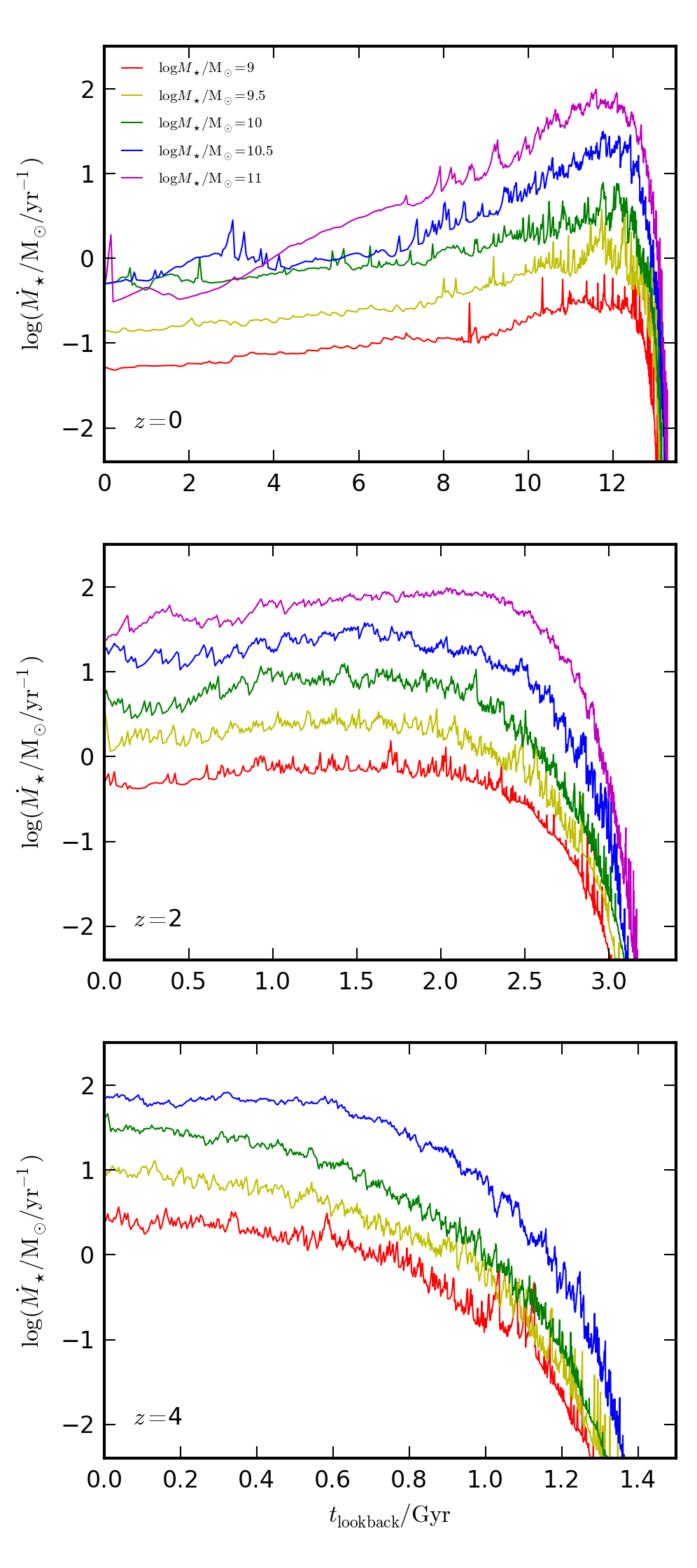}
\caption{Average SFHs of model galaxies in the Lagos12 model, plotted as a function of lookback time from the redshift labelled.
Each curve represents the average SFH of 100 galaxies of a given stellar mass at the redshift corresponding
to each panel, as indicated by the key.}
\label{sfh_evo}
\end{figure}

To explore this further, we show the average SFHs of galaxies from the Lagos12 model as a function of redshift and stellar mass in Fig.~\protect{\ref{sfh_evo}}. The average is performed over 
100 galaxies in each mass bin. It should be noted that for a large value of $\tau$, an exponential declining SFH resembles a constant SFH. With this in mind, it can be seen that, 
qualitatively, an exponentially decreasing SFH will provide an adequate fit to all of the average SFHs shown at lower redshifts. Even at $z=4$, the bulk of the stellar mass growth still 
occurs at a relatively constant star-formation rate such that an exponentially declining SFH fit to the data could recover the stellar mass if a large value of $\tau$ were chosen.

\begin{figure}
\includegraphics[width=20pc]{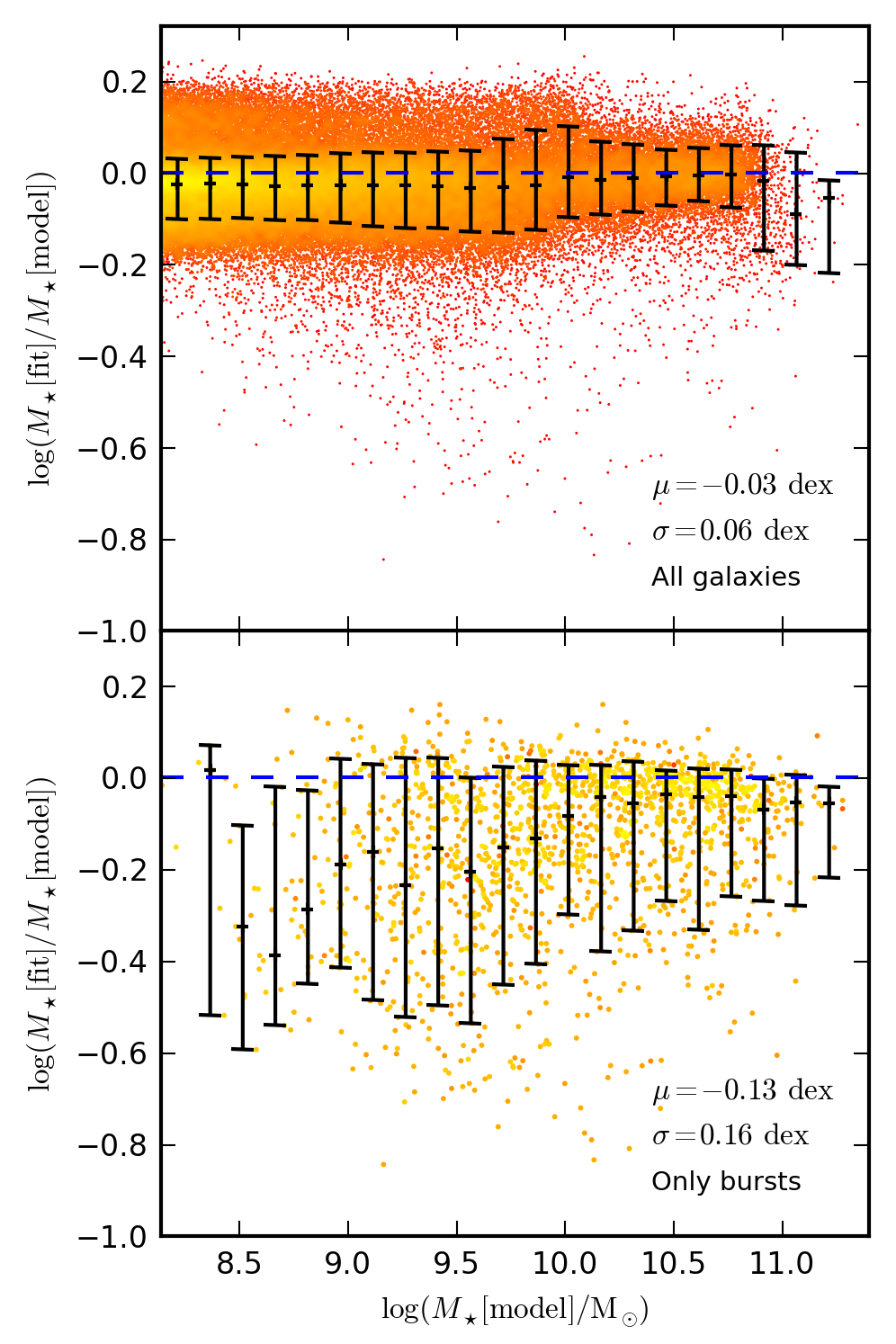}
\caption{The log of the ratio of the stellar mass estimated using SED fitting to the true stellar mass in the Lagos12 model at $z=2$, plotted as a function of the true stellar mass. 
As in Fig.~\protect{\ref{mass.recovery.flythrough.Lagos12}}a, both \galform and the SED fitting procedure have been modified such that all dust and metallicity effects are removed and the IMF, 
SPS model and treatment of recycling are consistent between the two calculations. 
The top panel shows the distribution for all galaxies.
The bottom panel shows the distribution for bursting galaxies, selected as model galaxies with a higher SFR in a burst component relative to the quiescent SFR in the galaxy disk. 
Formatting of points and symbols is the same as in Fig.~\protect{\ref{mass.recovery.evo.Lagos12}}.}
\label{mass.recovery.bursts.all}
\end{figure}

We strongly emphasize, however, that these findings only apply to the average over all model galaxies with $M_\star \geq 1.4 \times 10^8 \, \mathrm{M_\odot}$. The extremely small scatter
seen in Fig.~\ref{mass.recovery.flythrough.Lagos12}a. does not imply that the SFHs of individual galaxies are well recovered on an object-by-object basis. We have explored how well the SED 
fitting procedure recovers the mass-weighted mean age of model galaxies from the Lagos12 model and find a larger scatter (typically $\sigma \approx 0.2 \, \mathrm{dex}$) between estimated and true mass-weighted age than is shown 
for the mass recovery in Fig.~\ref{mass.recovery.flythrough.Lagos12}a.
Furthermore, a closer inspection of Fig.~\ref{mass.recovery.flythrough.Lagos12}a reveals that there are outliers to the overall distribution where the stellar mass is underestimated by almost an order of magnitude.
In Fig.~\ref{mass.recovery.bursts.all}, we demonstrate that if a subset of the overall galaxy population is considered, the assumption of an exponentially declining SFH can lead to larger 
errors in the stellar mass estimation. In this case we compare the average offset and scatter of the entire galaxy population from the Lagos12 model at $z=2$ with galaxies selected as being dominated 
by bursts of star-formation. We define bursts as galaxies with higher SFRs in a burst component relative to the SFR associated with quiescent star-formation in galaxy disks. We choose this subset 
because these galaxies are likely to have SFHs that differ significantly in many cases from an exponentially declining SFH. Our definition of a burst is a straightforward physical definition that 
can be made in a theoretical model, and should not be confused with the typical observational definition of a starburst as an object with an elevated SFR compared to the mean of the population. 
Comparison of the distributions shown in Fig.~\ref{mass.recovery.bursts.all} for all galaxies (top panel) and for bursts (bottom panel) shows that the stellar masses of bursting galaxies is underestimated
on average by $\mathrm{\Delta} \mu = - 0.1 \, \mathrm{dex}$ compared to the average over the total galaxy population. In addition there is significantly increased scatter in 
$M_\star\mathrm{[fit]}/M_\star\mathrm{[model]}$ when only bursts are considered.

Fig.~\ref{mass.recovery.flythrough.Lagos12}b shows a similar scenario to Fig.~\ref{mass.recovery.flythrough.Lagos12}a but where the SPS model, IMF and treatment of 
recycling are changed back to be consistent with the default Lagos12 model and default SED fitting procedure. Specifically, the Lagos12 model uses BC99 SPS models, 
instantaneous recycling and a Kennicutt IMF in this panel. The SED fitting procedure instead uses BC03 SPS models, non-instantaneous recycling and a Salpeter IMF. 
Comparison of Fig.~\ref{mass.recovery.flythrough.Lagos12}a and \ref{mass.recovery.flythrough.Lagos12}b shows that the different choices of SPS model and IMF, as well 
as the treatment of recycling, that can be made within SED fitting and \galform can cause constant offsets in $M_\star\mathrm{[fit]}/M_\star\mathrm{[model]}$ but do 
not create additional scatter in the distribution. It is interesting to explore the relative contribution from these different factors to these offsets. 
The remainder of Table~\ref{SPSimfRec} shows $\mu$ and $\sigma$ over a set of redshifts for different combinations of choices regarding recycling, the SPS model and the 
IMF. In all cases, dust and metallicity effects are removed from both \galform and the SED fitting procedure.

By default, the SED fitting procedure uses non-instantaneous recycling whereas \galform uses instantaneous recycling with a constant recycled fraction. This constant recycled fraction is 
fixed, for a given IMF, to the recycled fraction of a SSP with solar metallicity and age $10 \, \mathrm{Gyr}$. There are two factors that could lead to systematic errors in stellar mass 
estimation caused by differences between instantaneous and non-instantaneous recycling.
Firstly, the adopted relationship between initial and remnant mass for stars may be different in \galform and the BC03 SPS model used in the SED fitting procedure.
\footnote{In \galform, we use the relations between initial and remnant masses from \protect{\cite{Marigo96}} and \protect{\cite{Portinari98}}. See \protect{\cite{Cole00}} for details.}
We check this by comparing the recycled fraction at $10 \, \mathrm{Gyr}$ for a solar metallicity BC03 SSP with Salpeter IMF with the corresponding recycled fraction used by \galform for a Salpeter IMF. 
The two recycled fractions at $10 \, \mathrm{Gyr}$ are $R=0.31$ for the BC03 SSP and $R=0.30$ for \galform which are almost consistent. We therefore do not expect this factor to 
significantly affect the stellar mass estimation.
Secondly, for non-instantaneous recycling, the recycled fraction is a function of galaxy SFHs, whereas for instantaneous recycling the recycled fraction is independent of galaxy SFHs.
Therefore, as the overall age distribution of the model galaxy population evolves, it is to be expected that part of any systematic error in stellar mass estimation caused by differences 
between instantaneous and non-instantaneous recycling will be redshift dependent. This is verified by comparing the values of $\mu$ shown in the top and second sections of Table~\ref{SPSimfRec}. 
Changing from instantaneous recycling (top) to non-instantaneous recycling (second) in the SED fitting procedure has negligible impact at $z=0$ but results in a $15 \%$ offset in stellar mass 
by $z=4$. This is a small effect compared to some of the other potential sources of error (e.g. the choice of IMF) but should still be accounted for if an attempt is made to make a precise 
comparison between stellar masses derived from observations and theoretical models that use instantaneous recycling, particularly at high redshift. 

It is beyond the scope of this study to attempt to provide a comprehensive investigation into how stellar mass estimation is affected by uncertainties associated with SPS modelling and the 
IMF. Comparing the third and bottom sections of Table~\ref{SPSimfRec} shows that the difference between using a Salpeter and Kennicutt IMF is given by $\mathrm{\Delta} \mu \approx 0.25-0.29 \, \mathrm{dex}$
for $z \in 0,4$.
This simply demonstrates the well known result that changing from Salpeter to an IMF such as Kennicutt, that features a low-mass cutoff, results in $M/L$ ratios that are offset by nearly a constant factor, 
reflecting the fact that stars at the low-mass end contribute a negligible amount to the integrated light of a SSP. Comparing the second and third sections of Table~\ref{SPSimfRec} shows that there is 
also a small, $\approx 7\%$  increase in the scatter of the distribution at $z=0$ when the transition is made from using BC03 SPS in \galform (second) back to the BC99 SPS model (third) used in the default 
version of the Lagos12 model. 
We show this for reasons of completeness only because the BC99 and BC03 SPS models both belong to the same overall model family and are thought to be very similar. It 
should be noted that the difference between these two models almost certainly underestimates the true impact on stellar mass estimation associated with uncertainties in SPS modelling.

\subsection{Wavelength coverage}
\label{Wavelength}

\begin{table}
\begin{tabular}{|c|l|l|l|l|l|l|}

z & 0 & 0.5 & 1 & 2 & 3 & 4 \\
\hline
\multicolumn{7}{c}{All filters - Observer Frame} \\
$\mu /\mathrm{dex}$ & -0.01 & -0.01 & -0.03 & -0.02 & -0.01 & -0.01 \\
$\sigma /\mathrm{dex}$ & 0.01 & 0.03 & 0.04 & 0.06 & 0.05 & 0.05 \\
\hline
\multicolumn{7}{c}{All filters - Rest Frame} \\
$\mu /\mathrm{dex}$ & -0.01 & -0.01 & -0.01 & -0.01 & 0.01 & 0.0 \\
$\sigma /\mathrm{dex}$ &  0.01 & 0.02 & 0.03 & 0.02 & 0.02 & 0.03 \\
\hline
\multicolumn{7}{c}{No IRAC filters - Observer Frame} \\
$\mu /\mathrm{dex}$ & -0.01 & -0.01 & -0.03 & -0.04 & -0.03 & -0.03 \\
$\sigma /\mathrm{dex}$ & 0.01 & 0.03 & 0.05 & 0.11 & 0.10 & 0.10 \\
\hline
\multicolumn{7}{c}{No NIR or IRAC filters - Observer Frame} \\
$\mu /\mathrm{dex}$ & -0.01 & -0.02 & -0.04 & -0.05 & -0.07 & -0.07 \\
$\sigma /\mathrm{dex}$ & 0.02 & 0.04 & 0.09 & 0.19 & 0.29 & 0.23 \\

\end{tabular}
\caption{The mean median offset $\mu$ and half the mean $68\%$ range $\sigma$ of distributions in $M_\star\mathrm{[fit]}/M_\star\mathrm{[model]}$ against $M_\star\mathrm{[model]}$. 
As in Fig.~\ref{mass.recovery.flythrough.Lagos12}a, all values listed are for the Lagos12 model in the idealized case where dust and metallicity effects are removed and the choice of SPS model, 
IMF and treatment of recycling is consistent between \galform and the SED fitting procedure.
Each column corresponds to a different redshift. 
Each pair of rows corresponds to a different configuration of filters used to perform the SED fitting. 
The top pair of rows corresponds to the default case where the 12 broad-band, observer-frame filters listed in Table~\protect{\ref{SEDfittingParameters}} are used, spanning from $B_{435}$ to the 
$8.0 \mathrm{ \mu m}$ Spitzer IRAC band. 
The second pair of rows corresponds to the same filter set with the modification that the filters are fixed in the galaxy rest-frame. 
The third pair of rows corresponds to a reduced set of observer-frame filters where the Spitzer IRAC filters are removed. 
The final pair of rows extends this by removing the $J$, $H$ and $K$ bands along with the IRAC filters.}
\label{wavelength_coverage}
\end{table}

In the top section of Table~\ref{SPSimfRec}, we show the mean offset $\mu$ and mean spread $\sigma$ for a selection of redshifts 
in the idealized case where the Lagos12 model and the SED fitting routine are stripped back to the point where only the SFH 
is different between the SED calculations. For this idealized case, there is no systematic redshift dependence in the mean
offset $\mu$. However, there is a gradual increase in the scatter from $\sigma \approx 2\%$ at $z=0$ up to $\sigma \approx 
15\%$ at $z=2$. The scatter does not continue to increase beyond $z=2$. The increase in scatter over the redshift range 
$z \in 0, 1$ could be attributed to two separate effects. Firstly, any changes in the overall distribution of model galaxy 
SFHs with redshift could affect the accuracy of SED fitting. Over this redshift interval, there is a substantial fall with 
time in the overall SF activity, which may correspondingly reflect a change in the underlying distribution of SFHs. 
Secondly, the rest-frame wavelength coverage of the filter set changes with redshift such that the longer wavelengths in 
the rest-frame galaxy SED are no longer available in the SED fitting process at high redshift. To separate these two effects, 
we consider a modification of both the SED fitting procedure and \galform to use a filter set that is fixed in the galaxy rest-frame, 
independent of redshift. The top and second sections of Table~\ref{wavelength_coverage} show values of $\mu$ and mean spread $\sigma$ for 
the observer frame and rest-frame filter sets respectively. Using the rest-frame filter set removes most of the dependence 
of $\sigma$ on redshift, revealing that averaged over the entire galaxy population, any change in galaxy SFHs with redshift 
has a negligible impact on the accuracy of SED fitting when estimating stellar masses, at least when dust and chemical 
enrichment effects are not present.

For the sake of completeness, it is also interesting to explore how important the NIR filters are for accurately estimating 
stellar mass in this idealized case where dust, metallicity, SPS model and IMF related effects have all been removed. The 
bottom four rows of Table~\ref{wavelength_coverage} show $\mu$ and $\sigma$ in the case where either the IRAC filters or all of the NIR 
filters are removed from the SED fitting process. Comparison to the full observer-frame filter set shown in the top part 
of Table~\ref{wavelength_coverage} shows that having (perfect) photometry for the IRAC bands reduces the scatter in the stellar mass 
estimates by $\mathrm{\Delta} \sigma \approx 0.05 \, \mathrm{dex}$ for $z \geq 2$. In the scenario where only the optical 
filters are available, the accuracy of SED fitting degrades dramatically above $z=1$, even for the idealized scenario 
presented here. There is also an apparent trend whereby stellar masses are increasingly underestimated with increasing 
redshift. The degradation at higher redshifts demonstrates that it is necessary to sample the rest-frame optical-NIR part 
of the intrinsic galaxy SED in order to properly account for the contribution from older stars which typically dominate the 
stellar masses of galaxies.

\subsection{Metallicity}
\label{MetallicitySection}

\begin{figure}
\includegraphics[width= 20pc]{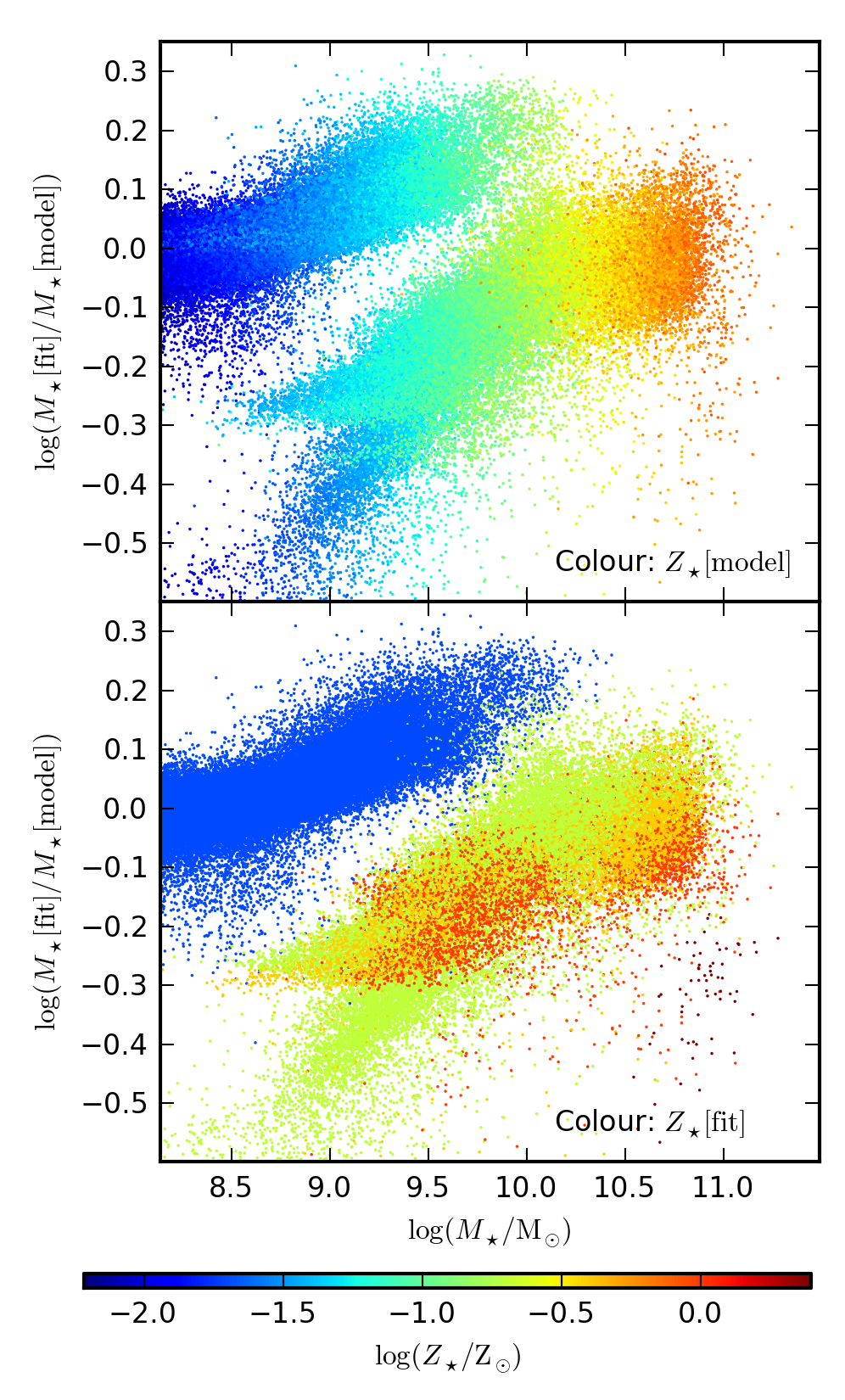}
\caption{The log of the ratio of the stellar mass estimated using SED fitting to the true stellar mass in the Lagos12 model at $z=2$, plotted as a function of the true stellar mass. 
As in  Fig.~\protect{\ref{mass.recovery.flythrough.Lagos12}}c, both \galform and the SED fitting procedure have been modified such that all dust effects are removed, and the IMF, 
SPS model and treatment of recycling are consistent between the two calculations.
{\it Top:} Each point represents an individual galaxy from the Lagos12 model and is coloured according the mean stellar mass-weighted metallicity calculated by \galform for that galaxy.
{\it Bottom:} Each point represents an individual galaxy from the Lagos12 model and is coloured according to the best-fitting metallicity solution calculated in the SED fitting procedure. 
The parameter grid in $Z_\star$ available to the SED fitting procedure is shown in Table~\protect{\ref{SEDfittingParameters}}}
\label{mass.recovery.Lagos12.showZ}
\end{figure}

Fig.~\ref{mass.recovery.flythrough.Lagos12}c reintroduces metallicity variation back into \galform and the SED fitting procedure.
For this panel, metallicity is a free parameter in the SED fitting procedure and the full chemical enrichment histories of model 
galaxies are used to calculate their SEDs in \galform. Fig.~\ref{mass.recovery.flythrough.Lagos12}c demonstrates that the bimodal 
features seen in Fig.~\ref{mass.recovery.evo.Lagos12} are caused by some aspect of the SED fitting calculation associated with 
metallicity. To better understand this behaviour, we show in the top and bottom panels of Fig.~\ref{mass.recovery.Lagos12.showZ}
the same distribution with galaxies colour coded by their mean mass-weighted metallicity in \galform or by the best-fitting 
metallicity calculated in the SED fitting procedure. This reveals that while the metallicity of galaxies in 
\galform is continuous across the bimodal feature, the metallicity returned by the SED fitting procedure clearly traces the 
bimodality seen in Fig.~\ref{mass.recovery.flythrough.Lagos12}c. This suggests that the SED fitting procedure could be 
incorrectly associating a metallicity with a model galaxy because of degeneracies with other parameters such as age 
\cite[e.g.][]{Bell01}. Underestimating the metallicity can lead to a corresponding overestimate of the age and hence the $M/L$ 
ratio. Additionally, it is possible that the parameter grid of metallicities used in SED fitting has insufficient resolution to 
reproduce the SEDs of model galaxies with mass-weighted metallicities that lie in between the values on the parameter grid.

\begin{figure*}
\begin{center}
\includegraphics[width=40pc]{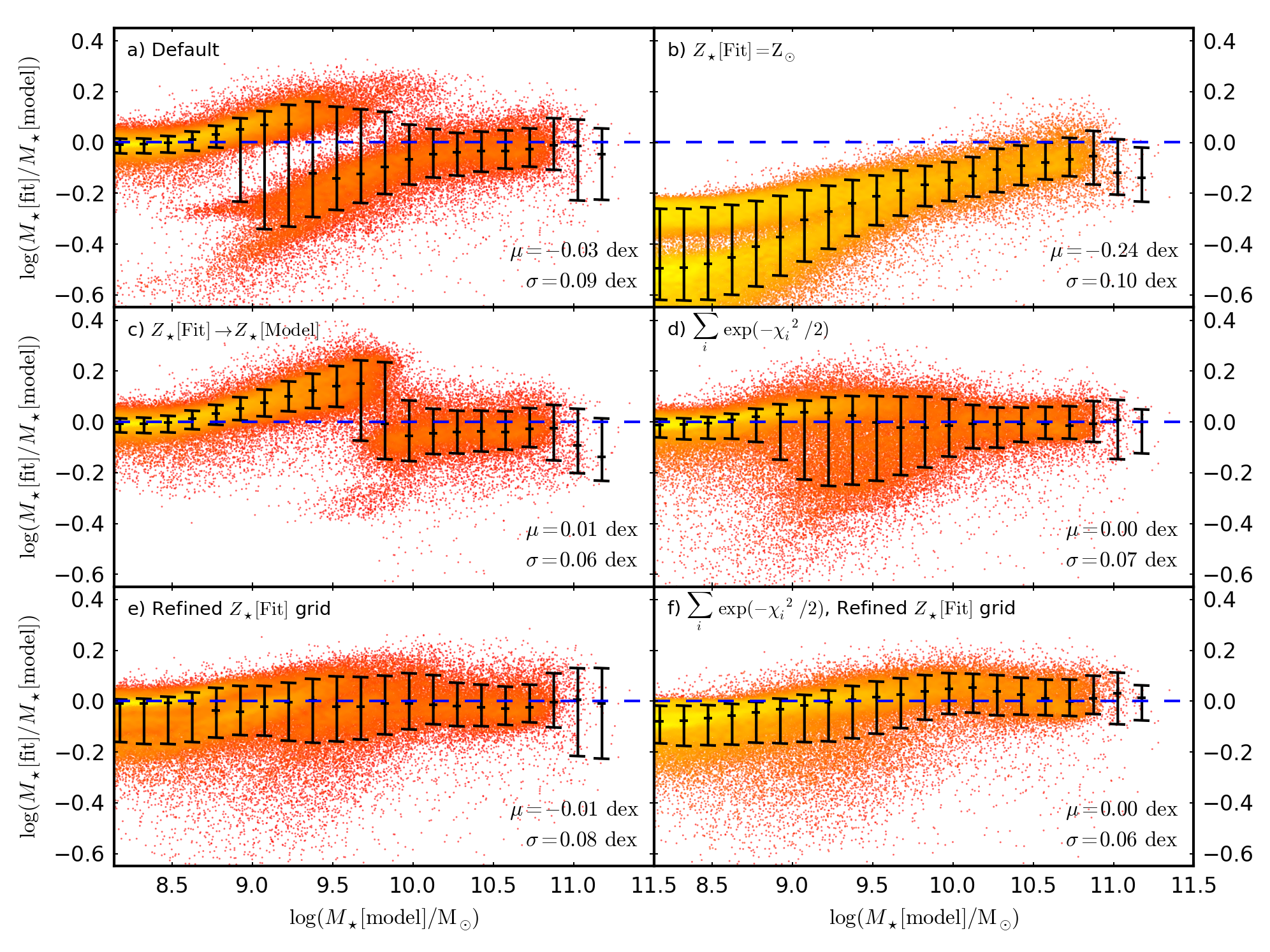}
\caption{The log of the ratio of the stellar mass mass estimated using SED fitting to the true stellar mass in the Lagos12 model at $z=2$, plotted as a function of the true stellar mass.
As in  Fig.~\protect{\ref{mass.recovery.flythrough.Lagos12}}c, both \galform and the SED fitting procedure have been modified such that all dust effects are removed and the IMF, 
SPS model and treatment of recycling are consistent between the two calculations.
Formatting of points and symbols is the same as in Fig.~\ref{mass.recovery.evo.Lagos12}.
Each panel corresponds to a different configuration of the SED fitting procedure.
{\it a)} The default case as shown in Fig.~\protect{\ref{mass.recovery.Lagos12.showZ}} where metallicity is a free parameter in the fit and the mode (best-fitting template) of the likelihood
distribution is used to estimate the $M/L$ ratio of each model galaxy.
{\it b)} Metallicity is constrained to $Z_\star = Z_\odot$ in the fit.
{\it c)} Metallicity is forced in the fit to use the closest possible value to the true mass-weighted metallicity of each model galaxy.
{\it d)} Metallicity is a free parameter in the fit and the mean over the likelihood distribution is used to estimate the $M/L$ ratio of each model galaxy. 
The mean is calculated using a likelihood-weighted summation $\sum_i \exp(-{\chi_i}^2 /2) M$ over the parameter space.
{\it e)} Metallicity is a free parameter in the fit and additional template SEDs are added to the template grid by interpolating in metallicity.
{\it f)} Metallicity is a free parameter in the fit, additional template SEDs are added to the template grid by interpolating in metallicity and the mean over the likelihood distribution is used to estimate the $M/L$ ratio of each model galaxy.}
\label{mass.recovery.Lagos12.ExtraZ}
\end{center}
\end{figure*}

In order to understand what is causing the bimodal behaviour and to see if it can be removed, we have explored a number of different choices regarding how metallicity
is treated in the SED fitting procedure. In Fig.~\ref{mass.recovery.Lagos12.ExtraZ}, we show how these choices affect the distribution of estimated to true stellar 
mass against stellar mass for the Lagos12 model at $z=2$, for the case where dust, recycling, SPS model and IMF related effects have been removed. The first and most simple option we explore is simply to fix the 
metallicity of all galaxies to a constant value in the SED fitting procedure. This choice is often made in observational studies presented in the literature \cite[e.g.][]{Rodighiero10,Marchesini09}. 
The distribution for this case is shown in Fig.~\ref{mass.recovery.Lagos12.ExtraZ}b. Although fixing the metallicity removes the bimodal behaviour, it also introduces a mass dependent 
bias into $M_\star\mathrm{[fit]}/M_\star\mathrm{[model]}$, whereby the stellar masses of less massive galaxies is underestimated. This behaviour is clearly 
undesirable, although the problem might be alleviated somewhat if a restricted range in stellar mass is considered, as is often the case for high 
redshift galaxy samples.

In Fig.~\ref{mass.recovery.Lagos12.ExtraZ}c, we force the SED fitting procedure to choose the closest metallicity on the parameter grid to the true mass-weighted metallicity of each model galaxy.
For each individual model galaxy, this is achieved by calculating the closest metallicity point on the template grid to the true mass-weighted metallicity and then excluding the other 
metallicity grid points as allowed solutions for that galaxy.
This choice could only be replicated in an observational study if external constraints were available on the stellar metallicity for each galaxy in the sample.
Comparison of the distribution shown in Fig.~\ref{mass.recovery.Lagos12.ExtraZ}c to the default case shown in Fig.~\ref{mass.recovery.Lagos12.ExtraZ}a shows that constraining the metallicity  in this way restricts
the bimodal behaviour to a narrow range in $M_\star\mathrm{[model]}$. This in turn indicates that there is a degeneracy between two possible metallicities which is 
broken when an external constraint is introduced. However, there is still a strong bimodality in the distribution at $M_\star \mathrm{[model]} \approx 4 \times 10^9 \mathrm{M_\odot}$.

Another choice that can be made in SED fitting is to change the statistical method used to obtain the best estimate stellar mass. Instead of picking the point in the parameter space 
with the smallest $\chi^2$ (which corresponds to the mode of the likelihood distribution), it is also possible to estimate the stellar mass of each galaxy by calculating the mean
over the likelihood distribution. This is achieved by performing a likelihood-weighted summation $\sum_i \exp(-{\chi_i}^2 /2)$ over the parameter space. \cite{Taylor11} describe
the implementation and advantages of this weighted-average method in more detail. In principle, taking the mean rather than the mode should result in estimated stellar masses that 
are more robust against discreteness in the parameter space and could therefore help to remove some of the bimodal behaviour seen in Fig.~\ref{mass.recovery.Lagos12.ExtraZ}a and 
Fig.~\ref{mass.recovery.Lagos12.ExtraZ}c. We show the distribution when this modified approach is used in Fig.~\ref{mass.recovery.Lagos12.ExtraZ}d. Comparison to 
Fig.~\ref{mass.recovery.Lagos12.ExtraZ}a reveals that the weighted-average approach does blur the bimodal feature, although the mass estimation is clearly still not perfect.

As a final step, it is also possible to simply interpolate template SEDs between the metallicity points on the SPS metallicity grid. This will help especially in situations where the 
likelihood distribution associated with a single metallicity grid point is significantly offset in $M/L$ ratio from the others. In this case, neither the mean nor mode of the distribution
will return a robust estimate of the true stellar mass if the outlying metallicity grid point dominates the overall distribution. We have confirmed that this is indeed the case for individual 
galaxies that fall on either side of the bimodal feature seen in Fig.~\ref{mass.recovery.Lagos12.ExtraZ}a. We add points to the parameter space at $Z_\star = 0.5, 0.3, 0.1, 0.05$ and 
$0.04 \, \mathrm{Z_\odot}$ using linear interpolation of the template SEDs in $\log(Z_\star)$. We show the effect of including this extended parameter grid in isolation in Fig.~\ref{mass.recovery.Lagos12.ExtraZ}e
and combined with the weighted-average method in Fig.~\ref{mass.recovery.Lagos12.ExtraZ}f. Adding the extra metallicity points in isolation breaks up the main bimodal feature into several 
smaller, less obvious features, improving the accuracy in the estimated stellar mass. Combining the interpolated metallicity grid with the weighted-average method almost completely 
removes any artificial bimodality in the distribution of the ratio of estimated to true stellar mass against stellar mass. The overall scatter is also slightly reduced in the process. 
Comparison with the scatter in Fig.~\ref{mass.recovery.flythrough.Lagos12}a shows that when metallicity is treated in this way, including metallicity in \galform and the fitting process does 
not adversely affect the stellar mass estimation.
This approach could easily be adopted in observational SED fitting. We discuss this further in Section~\ref{DiscussionMetallicity}.

\begin{figure}
\includegraphics[width=20pc]{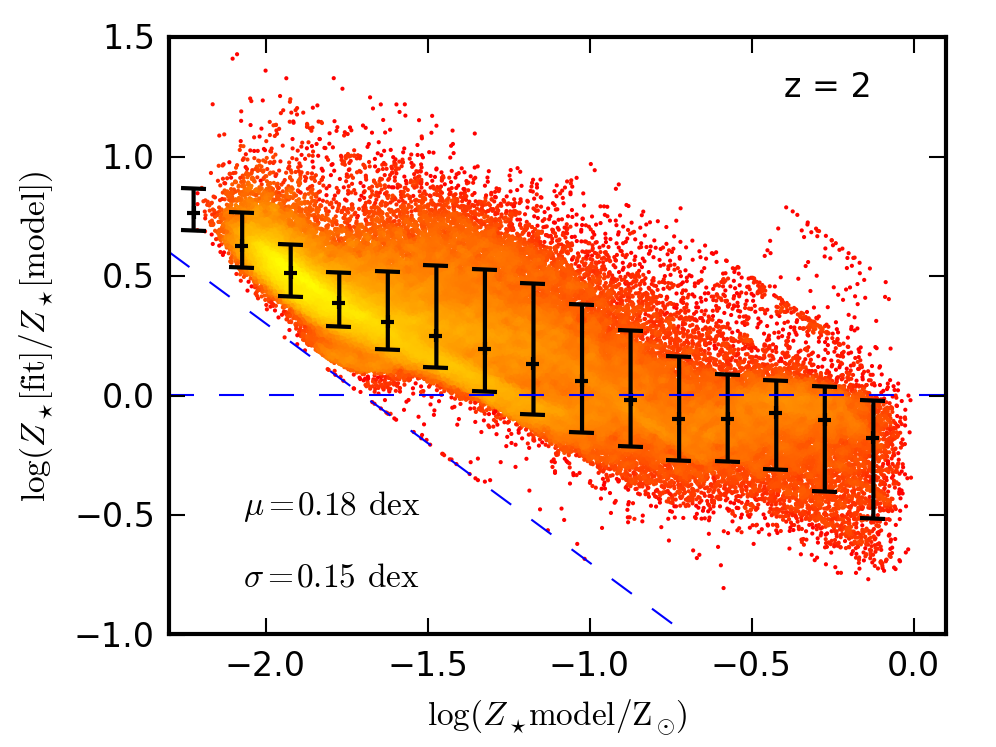}
\caption{The log of the ratio of the stellar metallicity estimated using SED fitting to the true mass-weighted stellar metallicity for model galaxies in the Lagos12 model at $z=2$, plotted as a function of the true mass-weighted stellar metallicity.
No dust effects are included in \galform and $E(B-V)=0$ is applied as a constraint in the fitting.
The Lagos12 model is modified to use BC03 SPS models with a Salpeter IMF and the SED fitting procedure is modified to use instantaneous recycling.
The SED fitting procedure has also been modified such that the best fit template is formed by a linear summation over all of the templates, weighted by the likelihood of each template. Also, additional templates are added through interpolation in $Z_\star$. 
The diagonal dashed line marks the lowest metallicity point on the SPS metallicity grid used in the SED fitting. Below this line, the SED fitting is not able to fit a metallicity that matches the metallicity in \galform.}
\label{Zrec}
\end{figure}

As an aside, it should be noted that the improved stellar mass recovery seen in Fig.~\ref{mass.recovery.Lagos12.ExtraZ}f does not directly imply that metallicity is also successfully recovered with
this modified fitting approach. In Fig.~\ref{Zrec}, we show the metallicity recovery corresponding to the distribution presented in Fig.~\ref{mass.recovery.Lagos12.ExtraZ}f. Comparing the two distributions 
shows that SED fitting is more successful at recovering stellar mass than metallicity ($\sigma = 0.15 \, \mathrm{dex}$ for metallicity and $\sigma = 0.06 \, \mathrm{dex}$ for stellar mass).

\subsection{Dust attenuation}
\label{dustresults}

\begin{figure}
\includegraphics[width=20pc]{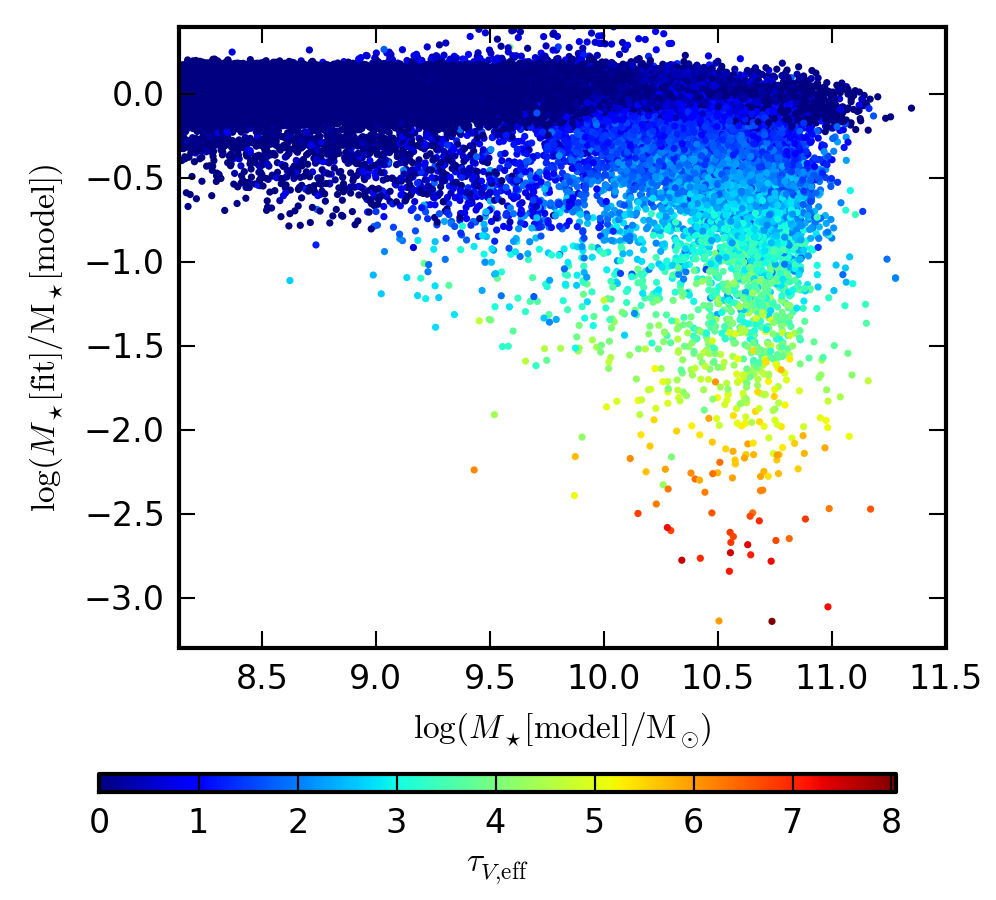}
\caption{The log of the ratio of the stellar mass estimated using SED fitting to the true stellar mass in the Lagos12 model at $z=2$, plotted as a function of the true stellar mass. 
As in  Fig.~\protect{\ref{mass.recovery.flythrough.Lagos12}}d, both \galform and the SED fitting procedure have been modified such that all metallicity effects are removed and the IMF, 
SPS model and treatment of recycling are consistent between the two calculations.
Each point represents an individual galaxy and is coloured by the rest-frame $V$-band effective optical depth, $\tau_{V\mathrm{,eff}}$, as calculated in the Lagos12 model.
The colour scaling is indicated by the key.}
\label{mass.recovery.Lagos12.dust}
\end{figure}

In Fig.~\ref{mass.recovery.flythrough.Lagos12}d, we reintroduce dust attenuation back into \galform and the SED fitting procedure. 
Fig.~\ref{mass.recovery.flythrough.Lagos12}d shows that some aspect related to how dust attenuation changes galaxy SEDs results in a 
population of galaxies which are intrinsically massive but have stellar masses which are significantly underestimated by SED fitting.
For less massive galaxies, reintroducing dust effects has a negligible impact on the stellar mass estimation because these galaxies
have small dust extinctions in the model. Furthermore, we find
that, averaged over the galaxy population, either including or excluding reddening in the fit (whilst retaining the dust extinction calculated by
\galform) actually has no impact on the recovered stellar mass.

\begin{figure*}
\begin{center}
\includegraphics[width=40pc]{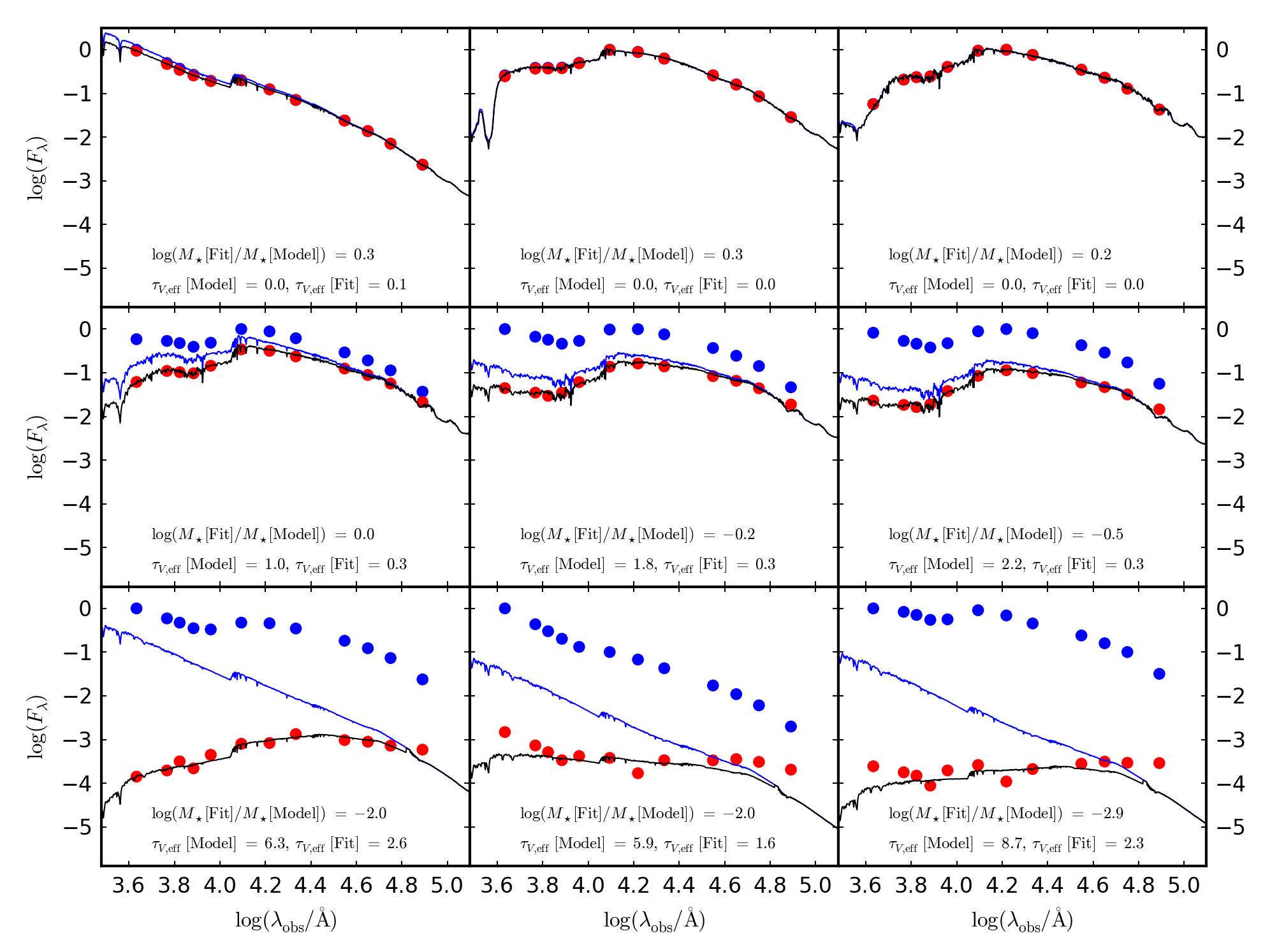}
\caption{SEDs plotted as a function of wavelength in the observer frame for 9 model galaxies generated by the Lagos12 model at $z=2$. 
The black and blue solid lines show the dust attenuated and intrinsic SEDs respectively for the best-fitting SED templates calculated by the SED fitting procedure. 
The blue and red filled points show the intrinsic and attenuated flux respectively for model galaxies in each of the 12 photometric bands used in the fitting process. 
For plotting purposes, the SEDs are normalized such that the maximum flux of each model galaxy, as calculated by \galform, is zero.
The three galaxy SEDs shown in the top row are selected quasi-randomly as examples where the SED fitting succeeds in recovering the intrinsic stellar mass.
The three galaxy SEDs shown in the bottom row are selected as those with the biggest mass offsets in $M_\star[\mathrm{Fit}]/M_\star[\mathrm{Model}]$. 
The remaining three galaxy SEDs shown in the middle row are intermediate cases between these two extremes.
$\log{M_\star [\mathrm{Fit}] / M_\star [\mathrm{Model}] }$ is the log of the ratio of the estimated to true stellar mass.
$\tau_{V\mathrm{,eff}} \mathrm{[model]}$ is the effective optical depth in the rest-frame $V$ band, as calculated in the Lagos12 model.
$\tau_{V\mathrm{,eff}} \mathrm{[fit]}$ is the effective optical depth in the rest-frame $V$ band, as estimated by the SED fitting procedure.}
\label{sed_worst}
\end{center}
\end{figure*}

\begin{figure}
\includegraphics[width=20pc]{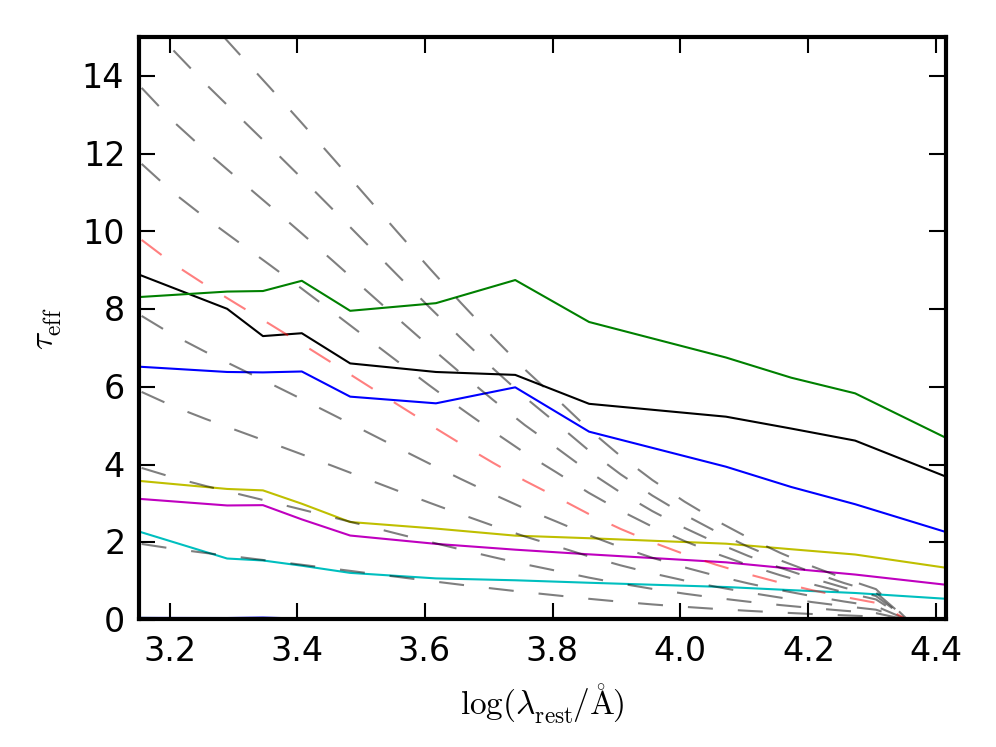}
\caption{Effective optical depth $\tau_{\mathrm{eff}}$, plotted as a function
of rest-frame wavelength at $z=2$. Solid coloured lines show the attenuation
of the Lagos12 model for the 6 galaxies shown in the middle and
bottom rows of Fig.~\protect{\ref{sed_worst}}. The dashed grey lines show the
attenuation curve from the Calzetti law for a range of values of $E(B-V)$.
The red dashed line corresponds to the specific case where $E(B-V)=1.0$, which is the
maximum reddening considered in the SED fitting procedure.}
\label{Lagos12.5mostdusty}
\end{figure}

Fig.~\ref{mass.recovery.Lagos12.dust} shows the same distribution shown in Fig.~\ref{mass.recovery.flythrough.Lagos12}d but 
with individual galaxies plotted as points coloured by their rest-frame effective optical depth in the $V$-band $\tau_{\mathrm{V,eff}}$, as 
calculated in the Lagos12 model. This shows a clear trend whereby the SED fitting procedure systematically underestimates the stellar masses of the 
model galaxies with the most dust extinction. Variations between the $M/L$ ratio of template SEDs with sensible combinations of parameters are typically much 
smaller than the offset in $M_\star\mathrm{[fit]}/M_\star\mathrm{[model]}$ seen for these highly extincted galaxies. This indicates that rather than the problem being caused
by parameter degeneracies between e.g. dust and age, it seems that the SED fitting is simply not correctly recovering the overall normalization of 
the intrinsic model galaxy SED. This implies in turn that the Calzetti law must be a poor match to the net attenuation curves calculated for dusty 
galaxies in \galform. We confirm that this is indeed the case in Fig.~\ref{sed_worst} and Fig.~\ref{Lagos12.5mostdusty}. Fig.~\ref{sed_worst} shows the intrinsic and
observed SEDs for 9 individual galaxies which are selected to show a range of offsets in $M_\star\mathrm{[fit]}/M_\star\mathrm{[model]}$. It is immediately apparent
that the SED fitting procedure underestimates the overall normalization of the intrinsic model galaxy SEDs for the dustiest galaxies. It is also
particularly noticeable that the radiative transfer calculation used in the Lagos12 model is applying a significant dust extinction to the entire 
SED, including up to the NIR for these galaxies. This behaviour cannot be reproduced by the Calzetti law. 

This can be seen more clearly in Fig.~\ref{Lagos12.5mostdusty}, which shows the attenuation curves for the 6 model galaxies shown in the middle and 
bottom rows of Fig.~\ref{sed_worst}. Also plotted is the Calzetti law for a wide range of values of $E(B-V)$. The red dashed line corresponds to the 
Calzetti law with $E(B-V) = 1$, the maximum value we include in the parameter space used in the SED fitting procedure.
Fig.~\ref{Lagos12.5mostdusty} demonstrates that although the Calzetti law can match the amount of attenuation applied to the dustiest galaxies by the
\galform radiative transfer model in the UV, it cannot reproduce the levels of attenuation at optical to NIR wavelengths. Furthermore, even if a different 
dust attenuation law was used in SED fitting that had the freedom to represent the types of attenuation curve of the model galaxies seen in 
Fig.~\ref{Lagos12.5mostdusty}, there would still be a very obvious degeneracy between the presence of a grey dust extinction component and simply having fewer stars 
producing light in a given galaxy. If such attenuation curves exist in reality, it would be very challenging to accurately constrain the stellar masses 
of dusty galaxies without performing detailed radiative transfer calculations using the entire UV - FIR SED.

In order to help understand why some of the dusty galaxies in the Lagos12 model have such extreme attenuation curves, it is also useful to consider the physical 
properties of the model galaxies shown in Fig.~\ref{sed_worst}. As discussed in Section~\ref{GALFORMSECTION}, galaxies are divided in \galform between a disk and bulge component.
In normal situations, dust is assumed only to be present in the disk but this changes during starbursts, triggered by mergers and disk instabilities. Starbursts
are assumed to take place inside the bulges of galaxies and therefore a bulge dust component is required to account for diffuse dust in these systems. We find that the 
dusty galaxies where the SED fitting procedure fails are typically compact and have high star formation rates. The most extreme examples shown in the bottom row
of Fig.~\ref{sed_worst} are compact starbursts. The intermediate cases shown in the middle row of Fig.~\ref{sed_worst} are compact, star-forming disks.
It is beyond the scope of this paper to provide a detailed analysis of the relationship between the physical properties of model galaxies from \galform and their dust attenuation
curves. However, it is straightforward to see from Equation~\ref{Eq.tauv0} that compact, gas rich galaxies will have the largest dust attenuations. In 
Section~\ref{DiscussionDust}, we discuss the origin of the shape of the attenuation curves shown in Fig.~\ref{Lagos12.5mostdusty}. We also discuss whether 
observed galaxies with high dust content could have similar dust attenuation curves.

\subsection{Alternative galaxy formation models}
\label{AlternativeSection}

\begin{figure}
\includegraphics[width=20pc]{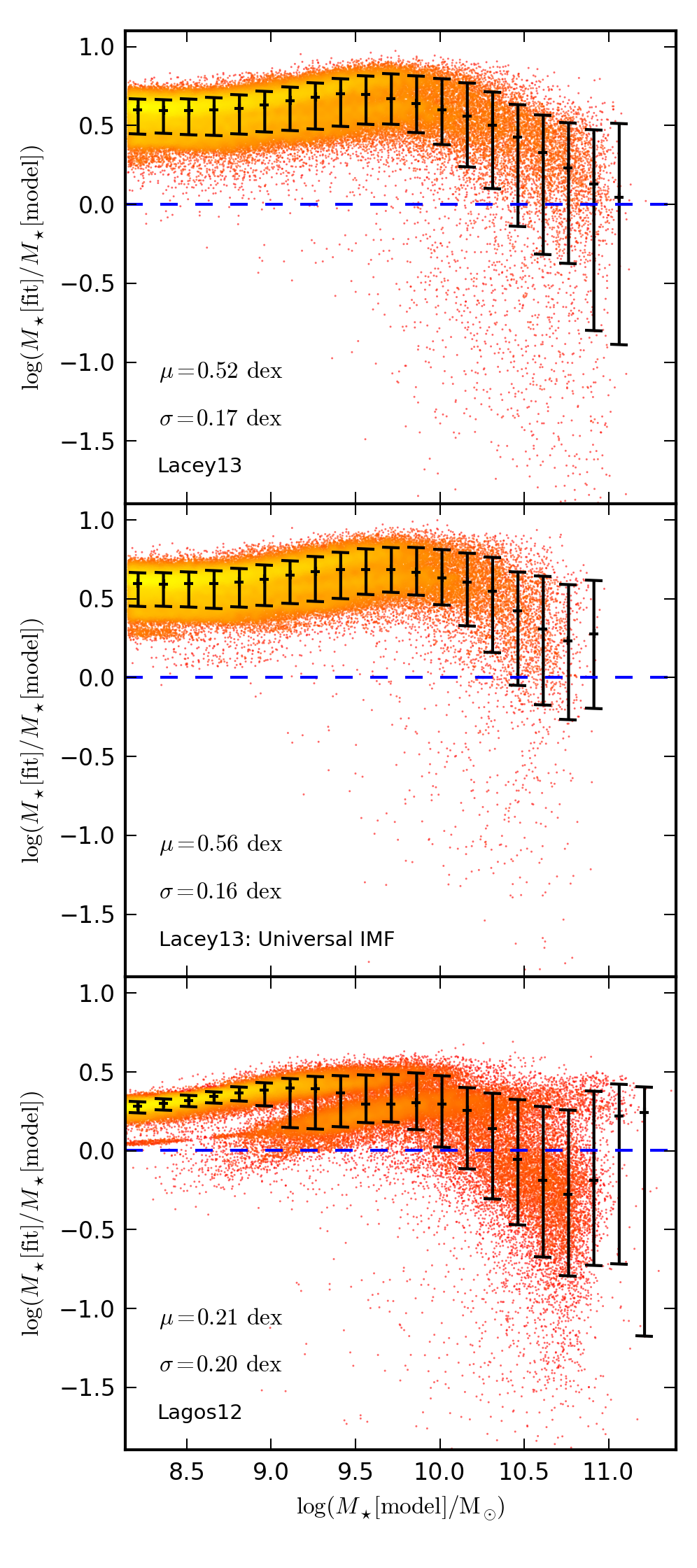}
\caption{The log of the ratio of the stellar mass estimated using SED fitting to the true stellar mass for different \galform models at $z=2$, plotted as a function of the true stellar mass. 
The top panel shows the distribution for model galaxies from the default Lacey13 model. 
The middle panel shows the distribution for model galaxies from a version of the Lacey13 model modified to use a universal Kennicutt IMF.
As a reference, the bottom panel shows the distribution for model galaxies from the Lagos12 model, as shown in \protect Fig.~\ref{mass.recovery.evo.Lagos12}. 
Formatting of points and symbols is the same as in Fig.~\protect{\ref{mass.recovery.evo.Lagos12}}. 
In all cases, the SED fitting uses a Salpeter IMF.}
\label{mass_recovery_model_comp}
\end{figure}

We explained in Section~\ref{Introduction} that the focus for this study is not to attempt to provide an exhaustive, quantitative guide on the accuracy of stellar
mass estimates derived from broad-band SED fitting. Part of the problem with using a SAM for this purpose is that results of this type will depend to some
extent on the specific choices and assumptions made as part of that SAM. Instead, we focus on a specific test case and try to understand and explain the origin of 
different systematics that appear in the distributions shown in Fig.~\ref{mass.recovery.evo.Lagos12}. As an extension of this analysis, we
also explore whether the behaviour seen in Fig.~\ref{mass.recovery.evo.Lagos12} is unique to the Lagos12 model. In Fig.~\ref{mass_recovery_model_comp}, we compare 
the distribution in estimated to true stellar mass against stellar mass from the Lagos12 model at $z=2$ with that in the Lacey13 model introduced in Section~\ref{GALFORMSECTION}. 
We use the same SED fitting procedure used to fit model galaxies from the Lagos12 model shown in Fig.~\ref{mass.recovery.evo.Lagos12}. The two distributions shown in the top and bottom panels
of Fig.~\ref{mass_recovery_model_comp} are quite similar in some respects but notably different in others. The general trend whereby the stellar masses of progressively 
more massive galaxies is increasingly underestimated is seen for both models. However, unlike for the results for Lagos12 model, the trend traced by the medians of the distribution 
continues monotonically all the way to the highest mass bins in the Lacey13 model, implying that, on average, even the very most massive galaxies in the Lacey13 model 
are very dusty. This implies in turn that they are forming stars at an elevated rate. This is an example of how our results can depend on the underlying physics that, 
in this case, controls the relative fraction of star-forming to passive galaxies.

The $\sigma$ values calculated for the two distributions show that there is a slightly smaller level of random errors when estimating the stellar masses of galaxies 
from the Lagos12 model, as compared to the Lacey13 model. Visually inspecting percentiles reveals that this difference can be attributed to the smaller scatter 
in the distribution in the low mass bins shown for the Lagos12 model. The $\Delta \mu \approx 0.3 \, \mathrm{dex}$ offset between the two distributions can be 
understood as a product of two separate effects. Firstly, the recycled fraction associated with the Kennicutt IMF used in the Lagos12 model is set to $R=0.39$. The 
corresponding recycled fraction used for stars forming in disks in the Lacey13 model is $R=0.44$. This can account for $\Delta \mu \approx 0.04 \, \mathrm{dex}$ 
of the total systematic offset. Secondly, the Lacey13 model uses the \cite{Maraston05} (MA05) SPS model to compute galaxy SEDs whereas both the Lagos12 model and the SED fitting procedure
use versions of the Bruzual \& Charlot SPS model family. As discussed in Section~\ref{GALFORMSECTION}, it is established that changing from using MA05 to BC03 SPS models 
in SED fitting of observed galaxies can change the estimated stellar masses by $\approx 50-60 \%$ \citep{Maraston06, Michalowski12}. If the TP-AGB contribution is really as uncertain
as the discrepancy between the BC03 and MA05 SPS models, the $\Delta \mu \approx 0.3 \, \mathrm{dex}$ offset between the two distributions has to be considered as a 
lower limit on the systematic uncertainty on stellar masses contributed by uncertainties from SPS modelling.

Combined with the uncertainty in the IMF, these differences actually result in a total offset of $\mu = 0.52$ for the Lacey13 model, relative to the SED fitting using a Salpeter IMF. 
As discussed in Section~\ref{GALFORMSECTION}, the Lacey13 model uses a top-heavy IMF of slope $x=1$ in starbursts. We have checked whether this is important for stellar mass estimation by performing SED fitting on
a modified version of the Lacey13 model that uses a universal Kennicutt IMF. The distribution for this scenario is shown in the middle panel of Fig.~\ref{mass_recovery_model_comp}.
We find that, averaged over the entire galaxy population, the distributions with and without the top-heavy IMF are very similar, at least at $z=2$. This is not unexpected because the optical-NIR SEDs 
(which to first order set the estimated stellar mass) of typical galaxies are unlikely, on average, to be dominated by light directly emitted by starbursting populations. However,
this will not necessarily be true for UV/FIR selected galaxy samples, particularly at higher redshifts. There is a small $\Delta \mu \approx 0.04 \, \mathrm{dex}$ mean offset
between the two distributions, whereby the estimated stellar mass is slightly higher, on average, for the universal IMF version of the Lacey13 model. Visual inspection of 
the percentiles of the two distributions shows that the systematic associated with dust, seen at the high mass end in Fig.~\ref{mass.recovery.flythrough.Lagos12}d, is slightly 
less prominent for this modified version of the Lacey13 model. The offset could therefore be caused by the higher rates of metal injection into the ISM that results from a 
top-heavy IMF in bursts. This in turn increases the dust content of the ISM in bursting systems, consequently increasing the impact of the systematics associated with dust.

Finally, the bimodal behaviour seen in $M_\star\mathrm{[fit]}/M_\star\mathrm{[model]}$ for the lower to intermediate mass bins in the Lagos12 model is not apparent for the Lacey13
model. We have investigated this further by performing SED fitting on the Lacey13 model in the case where dust effects are ignored. In this case, we find that
galaxies which are fitted with different metallicities are, on average, systematically separated in recovered stellar mass. This is in
agreement with the results seen in Fig.~\ref{mass.recovery.Lagos12.showZ} for the Lagos12 model. However, this effect is not visible in Fig.~\ref{mass_recovery_model_comp}
because any underlying bimodality in the distribution is blurred out by the larger overall scatter at low masses seen for the Lacey13 model. 
We have also explored how this situation changes at lower redshifts. At $z = 0.5$ the large offset caused by TP-AGB stars largely disappears. However, at this redshift, a similarly severe bias appears as a result
of metallicity discreteness effects. We find that metallicity grid effects at low redshifts and TP-AGB effects at higher redshifts mean that stellar masses are consistently overestimated by fitting model galaxies 
from the Lacey13 model for almost the entire redshift range considered in this study.

\section{The Stellar Mass Function}
\label{MassFunctionSection}

\begin{figure*}
\begin{center}
\includegraphics[width=40pc]{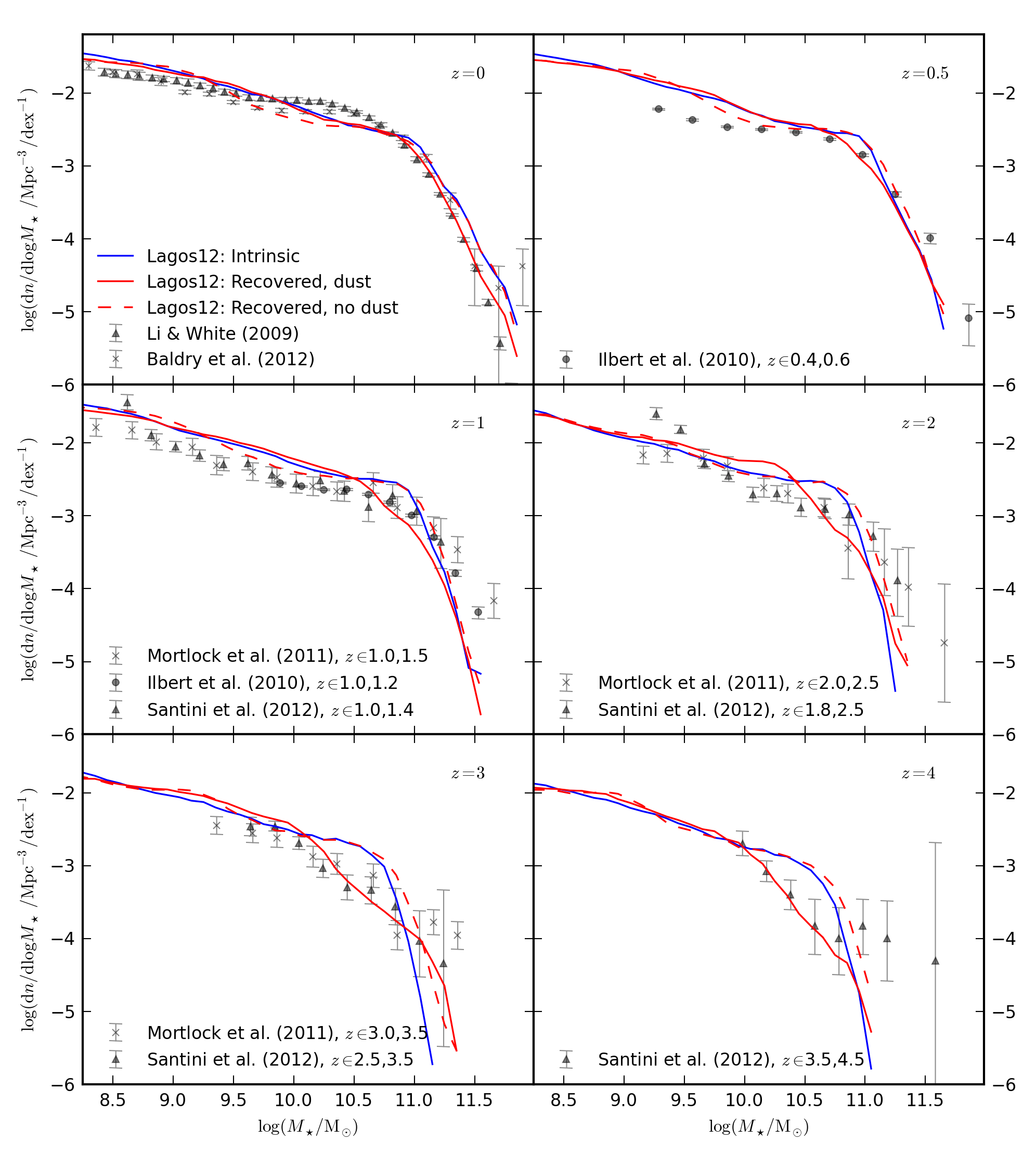}
\caption{Stellar mass functions predicted by the Lagos12 model for a selection of redshifts, as labelled in each panel.
The solid blue line shows the intrinsic stellar mass function produced by the Lagos12 model. 
The solid red line shows the stellar mass function recovered using SED fitting when dust effects are included and a Chabrier IMF is assumed in the fitting procedure. 
As a reference, the dashed red line shows the corresponding stellar mass function where no dust extinction is applied to the model galaxy SEDs and $E(B-V) = 0$ is used as a constraint in the fitting procedure. 
The grey points and error bars show observational estimates of the stellar mass function from \protect \cite{Li09}, \protect \cite{Baldry12}, \protect \cite{Ilbert10}, \protect \cite{Santini12} and \protect \cite{Mortlock11}. 
Where necessary we convert these observational results from a Salpeter to a Chabrier IMF using a $-0.24\,\mathrm{dex}$ correction, calculated by comparing the recovered stellar mass using Salpeter and Chabrier IMFs with
BC03 SPS models.}
\label{mf_evo.Lagos12}
\end{center}
\end{figure*}

In Section~\ref{MassRecoverySection}, we show that there are various systematics that can prevent the SED fitting procedure, described in Section~\ref{FITTINGSECTION},
from accurately estimating the stellar masses of individual model galaxies calculated by different \galform models. We now turn our attention to addressing the question of
how these systematics affect the global statistics of the galaxy population. We explore this issue by comparing the intrinsic stellar mass functions predicted by different 
\galform models to the corresponding mass functions recovered from the same models using SED fitting. Measurements of the stellar mass function are often used to constrain hierarchical galaxy 
formation models \cite[e.g.][]{Guo11, Henriques13}. It is therefore useful to also make a comparison with different observational estimates of the stellar mass function.
This also helps to put the systematic effects explored in Section~\ref{MassRecoverySection} into context. 

\begin{figure*}
\begin{center}
\includegraphics[width=40pc]{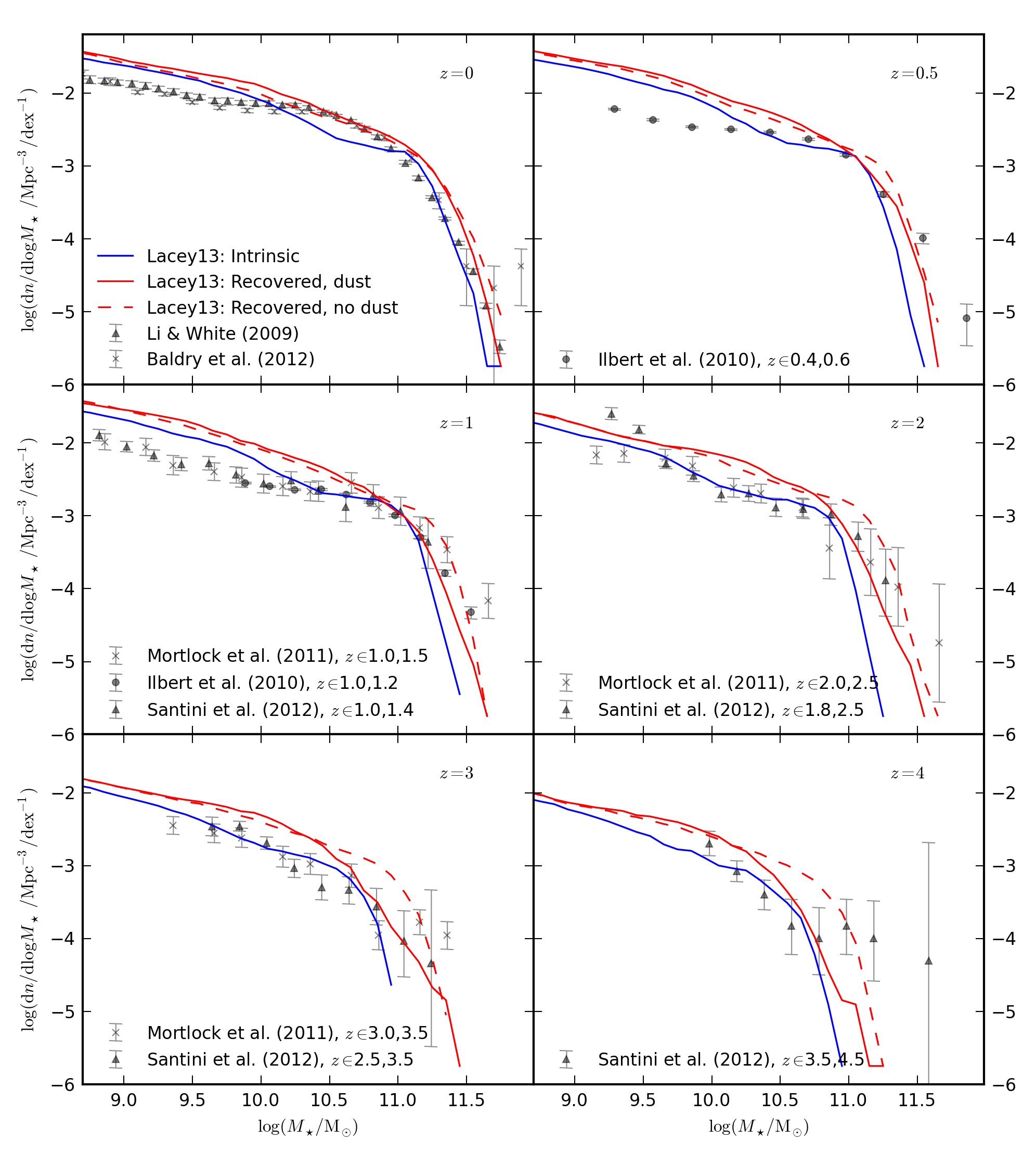}
\caption{Stellar mass functions predicted by the Lacey13 model for a selection of redshifts, as labelled in each panel. The definition and formatting of the lines and points is the same as in
Fig.~\protect{\ref{mf_evo.Lagos12}} }
\label{mf_evo.Lacey13}
\end{center}
\end{figure*}

We present stellar mass functions for a selection of redshifts 
from the Lagos12 model in Fig.~\ref{mf_evo.Lagos12} and from the Lacey13 model in Fig.~\ref{mf_evo.Lacey13}. We also present stellar mass functions for the \galform model
described in \cite{Baugh05} in an appendix.
We modify our standard SED fitting configuration (outlined in Table~\ref{SEDfittingParameters}) 
at this point by assuming a Chabrier IMF instead of a Salpeter IMF. This choice is made in order to be consistent with the bulk of the observational studies shown in Fig.~\ref{mf_evo.Lagos12} and 
Fig.~\ref{mf_evo.Lacey13}. For simplicity, we do not choose to change our SED fitting procedure between each redshift panel. It should be noted, however, that the different
observational studies all use slightly different variations of SED fitting parameters and filter sets. In addition, the low redshift observational studies 
\citep{Li09, Baldry12} use alternative SED fitting methods compared to the standard procedure described in Section~\ref{FITTINGSECTION}. \cite{Baldry12} use the likelihood-weighted summation technique, as explored
briefly in Section~\ref{MetallicitySection} and described in detail in \cite{Taylor11}. \cite{Li09} use stellar masses calculated with the nonnegative matrix factorization 
technique described in \cite{Blanton07}. For consistency with the results shown at other redshifts, we do not use these alternative methods for the mass functions recovered from the model.

Comparison of the intrinsic model mass function (solid-blue line) with the mass function recovered using SED fitting (solid-red line) in 
Fig.~\ref{mf_evo.Lagos12} shows that the systematics seen in Fig.~\ref{mass.recovery.evo.Lagos12} can have an appreciable impact on the
inferred global statistical properties of the galaxy population. The intrinsic and recovered model mass functions agree best at the low mass end
but disagree at the knee of the mass function and at the high mass end. This becomes increasingly evident in the higher redshift panels. 
The dominant factor responsible for this disagreement is dust, as discussed in Section~\ref{dustresults}. This is
demonstrated by comparing the recovered stellar mass functions when dust effects are (solid red line) and are not (dashed red line) 
included in the Lagos12 model and the SED fitting process. The recovered stellar mass function that includes dust extinction effectively
cuts away the knee of the intrinsic model mass function for $z > 0$. This is not seen when dust effects are not included, which can be
understood by comparing Fig.~\ref{mass.recovery.flythrough.Lagos12}a with Fig.~\ref{mass.recovery.flythrough.Lagos12}d. A common feature of the recovered
mass functions, both including and not including dust, is that in the highest redshift panels, the abundance of the most massive galaxies 
is increased with respect to the intrinsic stellar mass function predicted by the model. This simply reflects the Eddington bias where the exponential
decline of the mass function at the massive end means that any scatter in the stellar mass estimation shifts more galaxies into higher
mass bins than vice versa. This effect competes with the impact of dust attenuation to give the resultant shape of the solid red line in
Fig.~\ref{mf_evo.Lagos12}. Comparison to the observational data shows that, in some cases, the relative differences
between the intrinsic and recovered model stellar mass functions are much smaller than the disagreement with the observational estimates of
the stellar mass function, particularly at the low-mass end and at low redshift. In such cases, the model clearly does not accurately
reproduce the observational data and the stellar mass function can be used as a meaningful constraint. However, in the higher redshift panels the 
uncertainties on the observational data are larger and the differences between the intrinsic and recovered model stellar mass functions
also become larger. Taking the difference between the recovered and intrinsic model mass functions as a measure of the level of uncertainty 
regarding what model curve should actually be compared to the data, it is then apparent that it becomes difficult to place any meaningful 
constraints on the model using the observed stellar mass function at high redshift.

Fig.~\ref{mf_evo.Lacey13} shows the same information as Fig.~\ref{mf_evo.Lagos12} but for the Lacey13 model instead of the Lagos12 model.
The relationship between the intrinsic and recovered stellar mass function is similar to what is seen in Fig.~\ref{mf_evo.Lagos12} but 
there are also a number of differences. The most obvious difference is that the overall density normalization of the recovered stellar mass function 
is higher than for the intrinsic stellar mass function. This occurs predominantly because of differences between the MA05 SPS model used in the Lacey13 model 
compared to the BC03 SPS model used in the SED fitting procedure. As discussed in Section~\ref{AlternativeSection}, this means that the stellar masses of 
individual model galaxies are systematically overestimated for the Lacey13 model by SED fitting, shifting the overall stellar mass function to the right.
Comparison of the two recovered model stellar mass functions when dust effects are (solid red line) and are not (dashed red line) included shows
behaviour similar to what is seen in  Fig.~\ref{mf_evo.Lagos12}, whereby the perceived abundance of massive galaxies at and above the knee of the mass
function is suppressed, with the level of suppression increasing towards higher redshifts. It is particularly striking that the entire shape
of the mass function can change dramatically, even in the lower to intermediate redshift panels. At these redshifts, the intrinsic stellar mass
function predicted by the Lacey13 model does not resemble a single Schechter function as there is an apparent change in the power-law
slope before the break. This feature is washed out in the recovered stellar mass function, which instead resembles a single Schechter function 
when dust effects are included. Finally, as an aside, it is interesting to note that there is a fairly
strong level of disagreement between the shape of all of the model stellar mass functions and the shape of the observed mass function at $z=0$. 
A disagreement in the overall shape cannot be explained by any systematic uncertainty such as the $M/L$ associated with the IMF, and is interesting 
given that the model is tuned to reproduce the $z=0$ $K$-band luminosity function. This suggests that the stellar mass function can, at least at low
redshift, provide useful constraints for galaxy-formation models that are complementary to those provided by luminosity function data.

\subsection{Lyman-break galaxies}
\label{LBGsection}

\begin{figure}
\includegraphics[width=20pc]{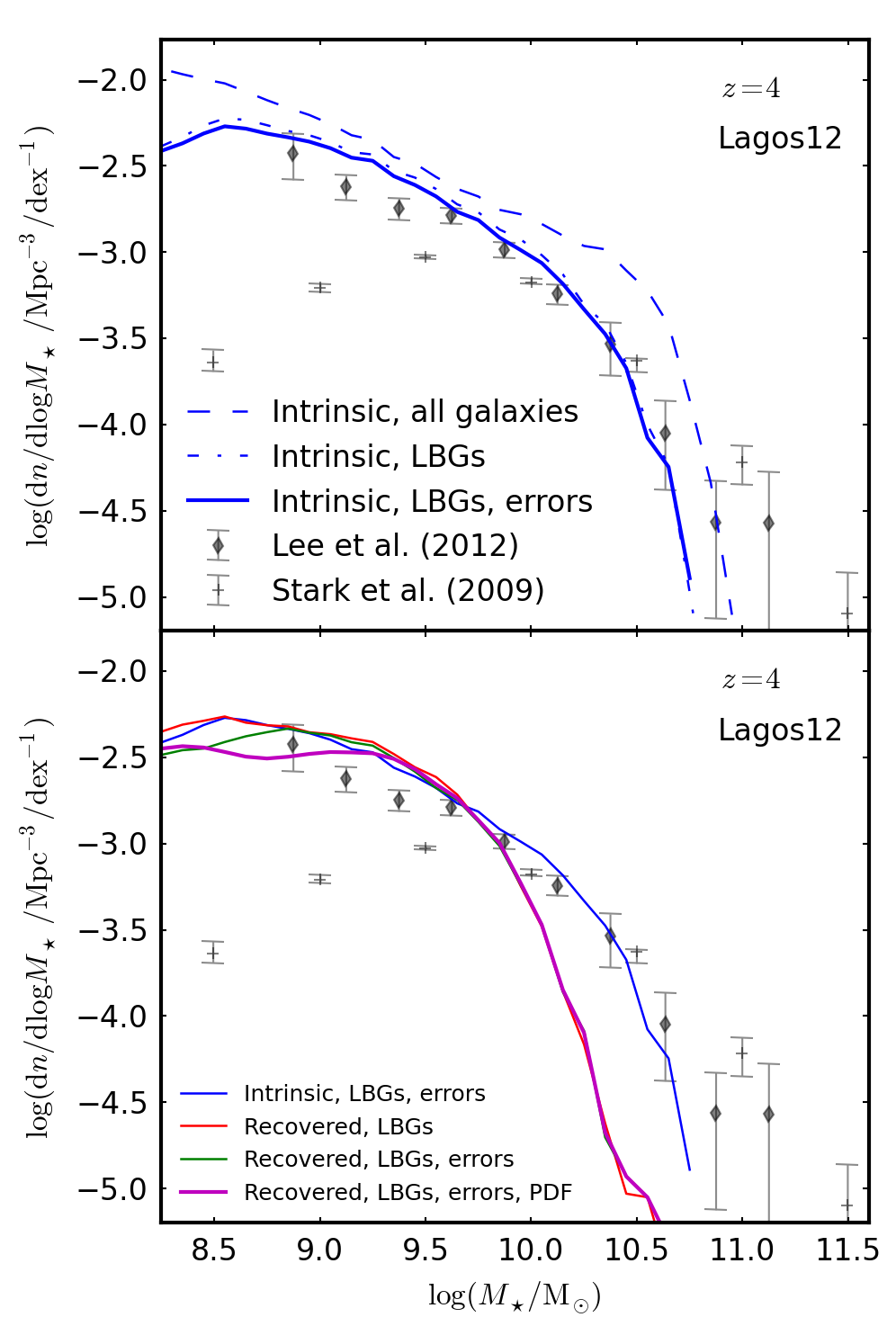}
\caption{Stellar mass functions predicted by the Lagos12 model at $z=4$. 
The top panel demonstrates how the intrinsic stellar mass function predicted by the model is reshaped as a result of LBG selection criteria.
The dashed blue line shows the intrinsic stellar mass function of all galaxies predicted by the Lagos12 model.
The dash-dotted blue line shows the intrinsic stellar mass function of galaxies selected using the $B$-dropout criterion from \protect\cite{Stark09} if no flux errors are included.
The solid blue lines, present in both panels, show the corresponding intrinsic stellar mass functions of LBGs when the selection includes the effects of artificial flux errors. 
The model galaxy fluxes are artificially perturbed to mimic the $S/N$ for each band quoted in Table $1$ of \protect\cite{Lee12}.
The bottom panel demonstrates how the stellar mass function of LBGs, as recovered using SED fitting, is reshaped as a result of errors on the photometry and the statistical method used to construct the mass function.
The red line shows the stellar mass function of LBGs recovered by SED fitting.
The green line shows the corresponding recovered mass function when the model galaxy fluxes are artificially perturbed. 
This line lies underneath the red and magenta lines at all but low masses.
The magenta line shows the corresponding recovered mass function when the model galaxy fluxes are artificially perturbed and the full stellar mass PDF of each individual galaxy is used to construct the mass function.
The points and corresponding error bars show the observationally inferred stellar mass functions from \protect\cite{Stark09} and \protect\cite{Lee12}. }
\label{mf_lbg_Lagos12}
\end{figure}

Up until this point we have focused on the relationship between the recovered and intrinsic stellar mass predicted for galaxies at low to intermediate 
redshift. At these redshifts, typically the photometric errors are small and colour selections do not need to be used in order to obtain well defined 
galaxy samples. Given that we find that the systematics in the distribution of $\mathrm{log}(M_\star\mathrm{[fit]}/M_\star\mathrm{[model]})$ can have an
appreciable impact on the recovered stellar mass function at these redshifts, it is interesting to explore how these effects translate to high redshift Lyman break galaxy (LBG)
samples, where the photometric and redshift errors can become large, colour selections exclude parts of the galaxy population and the available 
optical to NIR wavelength coverage in the rest frame is reduced.

We modify our SED fitting set-up at this point in order to more closely resemble the choices made by \cite{Stark09} and \cite{Lee12}, who estimate
the stellar mass function of LBGs. This alternative parameter grid is outlined in the bottom half of Table~\ref{SEDfittingParameters} and is described 
in Section~\ref{FITTINGSECTION}. We also consider a number of effects relevant for LBG samples that were not considered previously.
Firstly, we explore the effect of artificially perturbing the fluxes of model galaxies, following a Gaussian distribution consistent with the 
$5\sigma$ limiting depths listed in Table 1 of \cite{Lee12}. These limiting depths are also used to calculate $\sigma_n$ in Eq.~\ref{Eq.chi} when 
performing SED fitting. When the flux of a model galaxy falls below the $1\sigma$ limiting depth in a given band, we use the method of dealing with 
non-detections described in Section 3.1 of \cite{Lee12}.
Secondly, we explore the effect of LBG selection, taking the dropout criteria from \cite{Stark09} (both $S/N$ and colour selection criteria), which extends 
in redshift beyond \cite{Lee12} to include $z \approx 6$ $i_{775}$ dropouts. When exploring how the SED fitting performs without including the effects of
dust attenuation, we still use attenuated fluxes for the purposes of deciding which galaxies pass the LBG selection criteria and to decide which 
bands are counted as detections for individual galaxies. This ensures that the galaxy samples are consistent when comparing different recovered mass functions.
Finally, \cite{Lee12} construct their observed stellar mass functions by summing together the stellar mass probability distribution functions (PDFs)
calculated by their SED fitting procedure for each individual galaxy, hereafter referred to as the PDF method. This is distinct from standard practice in 
SED fitting where a single stellar mass value is assigned to each galaxy by considering only the best-fitting template to a given galaxy. This difference could 
potentially become significant at the low mass end of the mass function where galaxies are sufficiently faint that they are only detected in filters sampling 
the rest-frame UV. UV photometry does not provide strong constraints on the stellar mass associated with older stars in galaxies and this uncertainty will be accounted for on 
an object-by-object basis when using the PDF method to construct the stellar mass function. To provide a fair comparison to the mass functions from \cite{Lee12}, 
we also explore the effect of using the PDF method on our recovered mass functions.

In Fig.~\ref{mf_lbg_Lagos12}, we demonstrate how these various choices and optional modifications affect the stellar mass function predicted by the Lagos12 model at $z=4$.
The top panel shows how the intrinsic mass function predicted by the Lagos12 model is reshaped by LBG selection and artificial flux errors. LBG selection criteria
are designed to isolate UV-bright, star-forming galaxies over a given redshift range. For the Lagos12 model, this has the effect of reducing the normalization of 
the mass function in a fairly uniform manner over the range of masses considered here. The impact of including artificial flux errors has minimal impact on LBG
selection.
The bottom panel shows how the stellar mass function of LBGs, as recovered by SED fitting, is affected by artificial flux errors and the PDF method. All of
the recovered mass functions shown are virtually identical apart from at the low mass end. This simply reflects the fact that more massive galaxies are
typically brighter and are therefore detected with higher overall $S/N$ and at optical-NIR wavelengths where the photometry is not as deep.

\begin{figure*}
\begin{center}
\includegraphics[width=40pc]{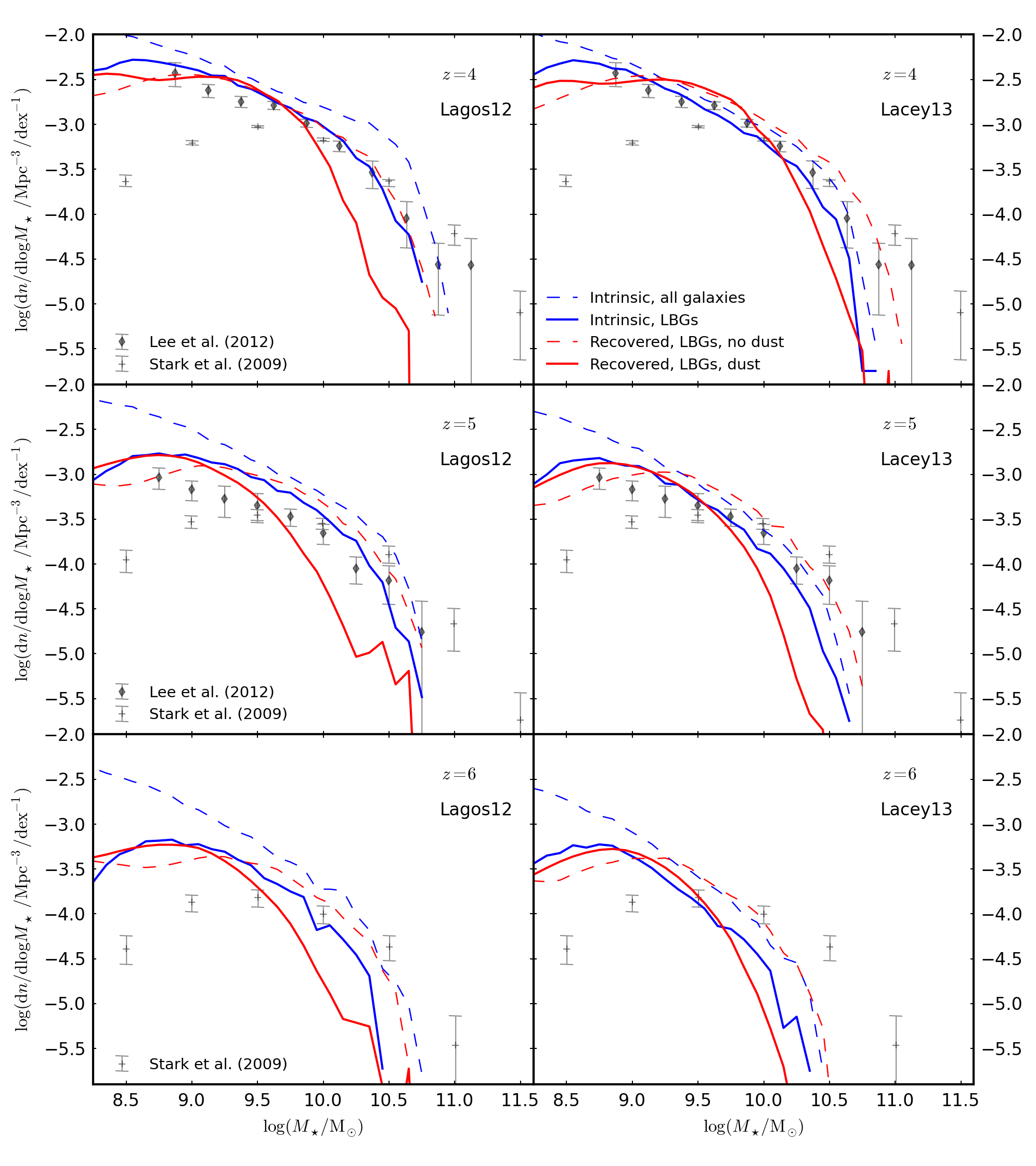}
\caption{Stellar mass functions predicted by the Lagos12 (left side) and Lacey13 (right side) models for a selection of redshifts, as labelled in each panel.
The dashed blue lines show the intrinsic mass functions predicted by the models without imposing any selection criteria.
In all of the other cases, model galaxy samples are constructed using the LBG selection criteria from \protect\cite{Stark09}. 
The flux of each galaxy is artificially perturbed to mimic the $S/N$ for each band quoted in Table 1 of \protect\cite{Lee12}. 
The solid blue lines show the intrinsic stellar mass functions of LBG-selected galaxies predicted by \galform. 
The solid red lines show the recovered model stellar mass functions of LBG-selected galaxies when dust effects are included in the SED fitting procedure. 
The dashed red lines show the corresponding recovered model stellar mass functions when dust effects are removed after selection criteria have been applied. 
All of the recovered mass functions are constructed using the full stellar mass PDF of each individual galaxy.
The points and corresponding error bars show the stellar mass functions from \protect\cite{Stark09} and \protect\cite{Lee12}. }
\label{mf_evo_lbg}
\end{center}
\end{figure*}

In Fig.~\ref{mf_evo_lbg}, we present recovered and intrinsic model stellar mass functions from the Lagos12 and Lacey13 models for a selection of redshifts.
All of the model mass functions shown are constructed using LBG selection criteria and artificially perturbed model galaxy fluxes. The PDF method is used to
construct the recovered mass functions, consistent with \cite{Lee12}. We also show measurements of the mass functions of LBG-selected samples from 
\cite{Stark09} and \cite{Lee12}. It should be noted that \cite{Stark09} apply an absolute magnitude cut at $M_{1500} = -20$, the effect of which can be 
clearly seen as the abundance of galaxies starts to fall below $M_\star \approx 2 \times 10^9 \mathrm{M_\odot}$. \cite{Stark09} also give their results for 
a Salpeter IMF which we correct by $-0.24 \, \mathrm{dex}$. This correction was estimated by comparing the recovered stellar mass using Salpeter and Chabrier 
IMFs with BC03 SPS models. When comparing the model mass functions with observational data, it should be noted that we do not attempt to mimic errors associated 
with photometric redshifts for our model galaxies. Including redshift errors is likely to increase the abundance of galaxies at the high-mass end, in line with 
observational samples that do not attempt to account for the Eddington bias associated with redshift errors.

On first examination, Fig.~\ref{mf_evo_lbg} shows a similar picture to that seen in Fig.~\ref{mf_evo.Lagos12} and Fig.~\ref{mf_evo.Lacey13}.
The inclusion of dust effects in the SED fitting procedure reshapes the recovered stellar mass function at the intermediate to high mass end (as seen by
comparing the solid and dashed red lines). It is interesting to see that while the intrinsic model stellar mass functions (blue lines) are quite a good 
match to the observed stellar mass functions in many cases, the model mass functions recovered using SED fitting are in very poor agreement when dust effects
are included. This emphasizes the danger of directly comparing intrinsic mass functions predicted by theoretical models to observational data at high redshift,
without accounting for the relevant uncertainties. It is also notable that the significant differences between recovered mass functions that include (solid red 
lines) and do not include dust effects (dashed red lines) are present in all of the panels shown. This demonstrates that for the models considered
in our analysis, dust continues to play a role in reshaping the entire UV-NIR SEDs of massive galaxies all the way out to $z=6$. Overall, similar to the situation 
seen in Fig.~\ref{mf_evo.Lagos12} and Fig.~\ref{mf_evo.Lacey13}, it is striking that the relationship between the recovered and intrinsic model stellar mass 
functions shown in Fig.~\ref{mf_evo_lbg} can vary dramatically, dependent on the different SED fitting choices which are made and the redshift considered.

\section{Discussion}
\label{DiscussionSection}

\subsection{The role of the SFH in estimating accurate stellar masses}
\label{DiscussionSFH}

Estimating the physical properties of galaxies from SED fitting requires the adoption of a prior distribution of galaxy SFHs. Large variations exist in the prior 
distributions used in different observational studies, reflecting the overall uncertainty in the optimal choice of SFH distribution.
As a consequence, considerable effort has gone into establishing how these choices
can affect the different galaxy properties that can be estimated from SED fitting \protect{\cite[e.g.][]{Lee09, Maraston10, Pforr12, Michalowski12, Banerji13, Schaerer13}}. \cite{Lee09} 
and \cite{Schaerer13} arrive at similar conclusions in that they both find that the stellar masses of LBGs are not particularly sensitive to the choice of SFH prior.
At face value, this seems to conflict with the conclusions from other studies which find that the estimated stellar masses of specific galaxy classes at high redshift are 
strongly sensitive to the assumed SFH distribution \cite[e.g.][]{Maraston10, Michalowski12, Banerji13}. It should be noted, however, that \cite{Michalowski12} and \cite{Banerji13} 
allow for the possibility of multi-component SFHs. This choice gives the SED fitting procedure more freedom to fit a very young, UV-bright stellar population at the same time as 
including a significant population of older stars that contribute to the total SED primarily at longer wavelengths. This approach is shown to yield systematically higher stellar 
masses relative to assuming a smooth exponentially declining SFH or an instantaneous starburst. 

We have demonstrated in the top section of Table~\ref{SPSimfRec} that, averaged over the entire galaxy population at a given redshift, the assumption of an exponentially 
declining SFH does not lead to a mean systematic offset in stellar mass. This result can be understood qualitatively by considering the shape of the average SFHs shown in Fig.~\ref{sfh_evo}.
In addition, we have shown that the scatter associated with fitting the SFH of model galaxies is negligible relative to the random and systematic errors caused by other factors in the SED modelling. 
We note that \cite{Lee09} also find that, averaged over a population of model LBGs, stellar masses are well constrained using an exponentially declining SFH.
It should be noted that this result only applies strictly to the case of fitting the SFHs of model galaxies in isolation, independent of dust and chemical enrichment effects. In practice, 
assuming a given prior SFH distribution may play a larger role in creating errors in stellar mass estimates because of the degeneracies that exist between the effects of age, metallicity 
and dust. Comparison of Fig.~\ref{mass.recovery.Lagos12.ExtraZ}f and Fig.~\ref{mass.recovery.flythrough.Lagos12}a shows that when dust effects are ignored and the treatment of metallicity 
in the fitting is improved, the result that assuming exponentially declining SFHs does not adversely affect stellar mass estimation still holds even when metallicity effects are reintroduced.
On the other hand, as an example of the problems caused by degeneracies associated with dust, \cite{Pforr12} show that in some cases it can be necessary to ignore dust entirely when applying 
SED fitting to galaxies with SEDs dominated by older stellar populations. This prevents the SED fitting procedure from incorrectly fitting young and highly reddened galaxy templates to these galaxies.

We have also demonstrated in Fig.~\ref{mass.recovery.bursts.all} that if galaxies are selected as being burst dominated, the errors on the estimated stellar masses associated
with assuming an exponentially declining SFH are greatly increased relative to the average error for the total galaxy population. This is not a particularly surprising result;
we have simply selected galaxies for which a single component, exponentially declining SFH is least likely to be appropriate. However, it does serve to reconcile our results with the findings of 
\cite{Michalowski12} in the sense that fitting smoothly varying SFHs to galaxies that are expected to have undergone recent bursts of star-formation (e.g. submillimeter galaxies)
can lead to systematically underestimating the stellar masses of these objects. 

\subsection{How should metallicity be included in SED fitting?}
\label{DiscussionMetallicity}

It is standard practice to either fix the metallicity or to use a small number of discretely spaced metallicities when performing broad-band SED fitting to estimate the physical 
properties of galaxies. Typically, interpolation between the metallicities is not used. This situation can be attributed to a combination of the inability of broad-band 
SED fitting to constrain metallicity \cite[e.g.][]{Pforr12}, the small number of SSP metallicities made available for popular SPS models and for reasons of numerical efficiency. Our 
analysis has shown that using a discrete and sparsely sampled metallicity grid causes undesirable bimodal features in the distribution of $\mathrm{log}(M_\star\mathrm{[fit]}/M_\star\mathrm{[model]})$, as seen
over a specific range in stellar mass in Fig.~\ref{mass.recovery.Lagos12.showZ}. We have also shown in Fig.~\ref{mass.recovery.Lagos12.ExtraZ}b that the decision to instead fix 
the metallicity of all galaxies to $Z_\star = \mathrm{Z_\odot}$ leads to an equally undesirable mass-dependent bias in $\mathrm{log}(M_\star\mathrm{[fit]}/M_\star\mathrm{[model]})$. 
This behaviour is straightforward to remove. Fig.~\ref{mass.recovery.Lagos12.ExtraZ}f demonstrates that using interpolation to add more metallicities to the parameter grid as well
as taking the mean rather than the mode of the probability distribution to estimate stellar mass can resolve any biases in stellar mass estimation associated with metallicity. This 
works primarily because interpolation acts to fill the gaps between metallicity grid points that are significantly offset in $M/L$ ratio for a given age but also because taking the mean helps to blur out discreteness in the distribution of $M/L$ ratios for a given parameter grid \cite[as discussed in ][]{Taylor11}. These steps could easily 
be incorporated into the standard SED fitting procedures used in observational studies.

The success of the revised fitting procedure shown in Fig.~\ref{mass.recovery.Lagos12.ExtraZ}f suggests that the standard assumption of a single stellar metallicity for all of the 
stars in a galaxy is acceptable, provided that interpolation and the method of performing a likelihood-weighted average over all templates is used. This is perhaps surprising given that 
model galaxies in \galform can have complex chemical enrichment histories. In particular, the metallicity of the individual stellar populations that make up model galaxies is strongly 
correlated with age. \cite{Conroy09} find that there is no significant difference between the optical and NIR colours of a single metallicity SSP and those of a multi metallicity SSP of the same average
metallicity. This would imply that SED fitting should correctly infer the $M/L$ ratios of galaxies, provided that the procedure selects the correct average metallicity for a given
galaxy. This is entirely consistent with our results. However, \cite{Conroy09} note that they do not account for a correlation between metallicity and age in their analysis which could
feasibly create a significant difference in the colours of single and multi-metallicity stellar populations. Given that such correlations exist for model galaxies 
in \galform, it is therefore reassuring that Fig.~\ref{mass.recovery.Lagos12.ExtraZ}f shows that this effect does not have a significant impact on the estimated stellar masses of galaxies. 
\cite{Gallazzi09} do find that fitting single or double colours of mock galaxies with chemical enrichment histories that contain an age-metallicity correlation has an impact
on inferred $M/L$ ratios. However, they find that this effect is small, consistent with our results. It should be noted that there is still a small level of mass-dependent bias evident in 
Fig.~\ref{mass.recovery.Lagos12.ExtraZ}f. Given the results of \cite{Gallazzi09}, this could potentially be explained as a result of fitting model galaxies' multi-metallicity stellar 
populations with single metallicity templates.

Stellar mass and stellar metallicity are extremely strongly correlated in the Lagos12 model, as can be seen in Fig.~\ref{mass.recovery.Lagos12.showZ}.
This is potentially significant in the context of our analysis because the strength of this correlation may serve to exaggerate the strength of the bimodal features seen in Fig.~\ref{mass.recovery.Lagos12.showZ}.
In addition, if the stellar metallicity of real galaxies at a fixed stellar mass differs from that of model galaxies predicted by the Lagos12 model, then the mass scale
where any bimodal behaviour appears will change, compared to the feature seen in Fig.~\ref{mass.recovery.Lagos12.showZ}. We also note that it is specifically the lowest sub-solar metallicities 
which are responsible for the offsets seen in  Fig.~\ref{mass.recovery.Lagos12.showZ}. If the stellar metallicities of real galaxies, at a given stellar mass, are higher than predicted
by the Lagos12 model, then it is possible that these sub-solar metallicities will not be relevant for the galaxies probed in most observational samples. In this case, the size
of the discreteness effects seen in Fig.~\ref{mass.recovery.Lagos12.showZ} will be significantly reduced. 

\subsection{Can galaxies have significant dust attenuation at optical to NIR wavelengths?}
\label{DiscussionDust}

The most significant source of error we encounter in estimating the stellar masses of model galaxies is found when 
SED fitting is applied to very dusty model galaxies. In Fig.~\ref{Lagos12.5mostdusty}, we have shown that these 
galaxies have much larger amounts of attenuation at longer wavelengths than is possible from the Calzetti law. 
This causes the stellar masses of dusty model galaxies to be significantly underestimated. In extreme cases, this underestimate
can be by factors as large as several hundred. From Fig.~\ref{mass.recovery.evo.Lagos12}, it can be seen that this 
error affects the majority of model galaxies above $10^{10} \, \mathrm{M_\odot}$ at $z > 2$. Consequently, for 
these redshifts, the errors associated with dust can completely reshape the recovered stellar mass functions shown 
in Fig.~\ref{mf_evo.Lagos12}. Accordingly, the intrinsic rest-frame optical and NIR luminosity functions predicted 
by \galform will also be reshaped by dust. This will have a significant impact on any attempt to compare this 
particular model with observational data at higher redshifts. 

In order to understand this overall result, it is useful to consider how there can be significant dust attenuation 
at optical to NIR wavelengths for model galaxies in \galform. Firstly, it is important to note that for the 
assumption of a star-dust geometry corresponding to a uniform foreground dust screen, and for a specific dust grain model,
the ratio of absolute extinction $A_{\mathrm{V}}$ relative to the reddening $E(B-V)$ must be constant. \cite{Calzetti00} assumed this 
star-dust geometry and applied energy balance arguments to UV and FIR observations of 4 local starbursts to fix 
$R_{\mathrm{V}} = A_{\mathrm{V}} / E(B-V) = 4.05$. Once $R_{\mathrm{V}}$ has been fixed in this way, the 
attenuation curve from \cite{Calzetti94} can no longer reproduce the attenuation curves shown in 
Fig.~\ref{Lagos12.5mostdusty}. 

In reality, the assumption of a uniform dust screen is a very poor approximation to a realistic star-dust geometry. 
In \galform, disk stars are embedded in a diffuse dust component with the same spatial distribution as the stars. 
In this case, the path length through the diffuse dust to the observer will be different for each star. Therefore, 
the light from each star will experience a different amount of attenuation, yielding a net attenuation curve for 
the entire galaxy which can be significantly different from the input extinction curve of the dust grains \citep{Gonzalez-Perez13}. A 
simple example that demonstrates this behaviour is to consider a star-dust geometry corresponding to an infinite 
uniform slab containing stars and dust mixed together with the same uniform spatial distribution. The effective 
optical depth, $\tau_{\mathrm{eff}}$, for this geometrical configuration is given by

\begin{equation}
\tau_{\mathrm{eff},\lambda} = - \ln{\left( \frac{1 - \exp{( - \large{\tau_{0,\lambda}} \sec{i}} )}{ \large{\tau_{0,\lambda}} \sec{i}} \right)},
\label{Eq.InfSlab}
\end{equation}

\noindent where $i$ is the inclination angle of the slab relative to the observer and $\tau_{0,\lambda}$ is the 
face-on extinction optical depth for a single sightline through the slab. In the limit that $\tau_{0,\lambda}$ 
becomes large, so that the slab is optically thick, this simplifies to 
$\tau_{\mathrm{eff},\lambda} \simeq \ln(\tau_{0,\lambda} \sec{i})$. In this scenario, nearly all of the light 
emitted by stellar populations from within the slab is absorbed and only light from a layer of stars at the surface 
can reach the observer. The light that escapes the slab only passes through a small amount of diffuse dust and as a 
consequence, is only reddened by a small amount. Therefore, for the optically thick case, this configuration will 
yield net attenuation curves which, compared to the Calzetti law, are considerably greyer. This explains how the 
attenuation curves of very dusty galaxies in \galform can have significant amounts of attenuation at long 
wavelengths.

In reality, dust in galaxies is not thought to exactly trace the spatial distribution of stars in galaxy disks. By 
comparing the reddening of nebula emission (i.e. through the $H_\alpha$ to $H_\beta$ line luminosity ratio) with the reddening observed in the total UV 
continuum of galaxies, it is apparent that dust in the ISM must be concentrated around star-forming 
regions \cite[e.g.][]{Calzetti94}. This observation has motivated the use of two-component dust models 
\cite[e.g.][]{Silva98} that contain a compact, birth-cloud dust component that attenuates the light emitted by very 
young stellar populations. This scheme is applied in \galform and as such, the net attenuation curves of model galaxies 
shown in this paper take this effect into account. However, the diffuse dust in galaxies is also not thought to exactly 
trace the spatial distribution of stars. Specifically, the scale height of diffuse dust is known to be smaller than the 
overall scale height of stars in galaxy disks, particularly when considering older stellar populations 
\cite[e.g.][]{Wild11}. 

Additionally, the presence of a clumpy ISM means that there are likely to be some sightlines 
through galaxy disks which are relatively free of dust. This effect was considered by \cite{Conroy10a} who explored the
effect of a clumpy ISM with a lognormal column density distribution of dust combined with the empirical dust model from 
\cite{Charlot00}. Although they show that the clumpiness of the ISM (characterized by the width of the lognormal column 
density distribution) can have a large impact upon NUV and optical colours, they state that there is negligible impact
from dust on the $K$-band luminosity function. It should be noted that this conclusion depends strongly on how they 
calculate the total amount of dust in the diffuse ISM. If there is a sufficiently large mass in diffuse dust, a clumpy 
ISM will produce similar emergent behaviour in terms of the shape of the total attenuation curves to that of the slab 
geometry considered earlier. That is, the total SED of a dusty galaxy with a clumpy ISM will be dominated by light 
emitted by stars that lie along relatively unobscured sightlines (which experience only minimal reddening) while light 
emitted from behind or within optically thick regions will be completely absorbed in comparison. However, the effect of 
having a clumpy ISM will, for an equal mass in diffuse dust, result in a lower normalization of the total attenuation 
curve as compared to an optically thick, uniform slab. This is simply because a higher fraction of the stars will be 
unobscured for the case of a clumpy ISM. At present, the dust modelling in \galform does not account for any clumpiness 
of diffuse dust in the ISM. This may result in a higher normalization for the attenuation curves of very dusty model 
galaxies, compared to real dusty galaxies. Although the inclusion of a birth-cloud dust component does account in part 
for clumpiness in the ISM, the impact from birth clouds will be of secondary importance if the diffuse dust component 
contains enough mass to absorb all of the light not emitted from close to the surface of the galaxy disk. 

The problem of choosing an appropriate star-dust geometry becomes even more complex in the case of a galaxy merger. 
It is well established from numerical simulations that pressure forces experienced by gas in the ISM can decouple the 
spatial distributions of gas relative to stars, such that gas is funnelled into a compact region in the centre of the 
system, producing a nuclear burst of star-formation. \cite{Wuyts09} show that by applying SED fitting to a suite of 
idealized hydrodynamical simulations of galaxy mergers, the stellar masses of simulated galaxies can be systematically 
underestimated. This occurs because as the diffuse dust is concentrated into the central region, any light emitted from 
within or behind that region will be almost entirely cut out of the observed galaxy SED. In addition, the overall 
stellar distribution will be much more spatially extended than the gas during this phase and consequently will suffer 
minimal reddening. This is exactly analogous to the behaviour discussed previously for an embedded star-dust geometry 
or a clumpy ISM. However, as noted by \cite{Wuyts09}, the situation is complicated in this case by the fact that the 
stellar populations which are heavily obscured are younger, on average, compared to the total stellar population. A 
strong correlation between stellar population age and the dust column density will serve to dilute the greying effect
discussed for the slab and clumpy ISM examples \cite[see Section 4.3.2 in ][]{Wuyts09}. 
At present, \galform fails to account for these effects in specific 
situations; in the event of a major merger or disk instability, stars of all ages are mixed evenly with diffuse dust 
and gas in the galaxy bulge. This geometry is unlikely to be representative of real merging systems (although the 
situation is far less clear at high redshift) and as such, the attenuation curves of systems in \galform that are 
undergoing major mergers or disk instabilities may be unrealistic. In this case, the impact of dust on stellar mass 
estimation could be exaggerated to some extent for these systems. Changing the radial scale length of the burst 
compared to the stellar bulge would represent only a small change to the current implementation in \galform and we plan 
to investigate the impact of this change in future work.

Aside from uncertainties associated with the star-dust geometry, it is important to appreciate the uncertainties 
associated with calculating the mass and density of dust in the ISM of galaxies, particularly at high redshift. 
Generally speaking, theoretical galaxy formation models that attempt to model dust use local relations that give the 
ratio of dust to metals in the ISM \citep{Cole00}. These local relations are then applied universally, which is a 
large extrapolation in the case of actively star-forming galaxies at high redshift where the physical conditions in
the ISM can be very different. In addition, other aspects of a given theoretical galaxy formation model will have a 
strong impact on the final effect of dust on model galaxy SEDs. For example, in the case of our analysis, if the 
calculations of metallicities or galaxy sizes are incorrect in \galform, then the size of the errors associated with 
dust on stellar mass estimation will also be incorrect. 

In summary, the assumption of a uniform foreground dust screen must be incorrect in most cases for real galaxies, but 
the details of the star-dust geometry in real galaxies may also be more complex than what is assumed in \galform. In 
addition, both the mass and density of dust calculated by \galform is dependent on both the overall accuracy of the 
model and being able to extrapolate the local dust to metal ratio up to high redshift. We plan to explore how these 
factors could affect our results for stellar mass estimation in future work.

Having explained why dusty galaxies in \galform can have significant amounts of attenuation at longer wavelengths 
relative to the Calzetti law, it is useful to consider if there is any evidence for this behaviour from other 
observational or theoretical studies. \cite{Pforr12} and \cite{Lee09} do not find any evidence that dust can cause 
the stellar masses of galaxies to be significantly underestimated when they fit model galaxies from other SAMs. 
However, we emphasize that the SAMs considered in these studies do not attempt to model dust attenuation in a 
physically motivated way. Instead, they adopt the same Calzetti attenuation curve in the SAM as in the SED fitting.
It is therefore of no surprise that these authors do not recover the same results shown by our analysis. 

\cite{LoFaro13} fit the full UV-FIR SEDs of an observed sample of 31 luminous and ultraluminous infrared galaxies at 
$z=1$ and $z=2$. From their Figures 2, 5 and 7 it can be seen that their fitting method estimates significant 
attenuation of the stellar continuum across the entire UV-NIR SED for a number of objects in their sample. They also 
apply a standard UV-NIR SED fitting procedure to the same galaxy sample and find that, for the most dust obscured 
galaxies, the stellar mass obtained in this case is underestimated relative to what is estimated from the full 
UV-FIR fitting procedure. In the top-left panel of their Figure 6, they show a clear trend whereby stellar mass is 
increasingly underestimated by UV-NIR SED fitting for increasingly dust obscured systems. This is in qualitative 
agreement with our results. It should be noted that they use the radiative transfer code \grasil \citep{Silva98} to 
generate UV-FIR templates. \grasil assumes the same star-dust geometry as is assumed in \galform. On the one hand, 
this assumed geometry is physically motivated and is clearly superior to the crude assumption of a uniform 
foreground dust screen. However, the various uncertainties discussed earlier associated with choosing this particular 
star-dust geometry are also relevant for their analysis. 

\cite{Michalowski10} also used \grasil to fit the full UV-FIR SEDs of a sample of 76 spectroscopically confirmed 
submillimeter galaxies, estimating stellar masses for these objects. \cite{Michalowski12} then revisited the same 
sample but instead applied a standard UV-NIR SED fitting procedure to estimate stellar masses. As discussed in 
Section~\ref{DiscussionSFH}, the intention behind their analysis was to investigate how priors on the SFH 
distribution of submillimeter galaxies can affect stellar mass estimation for this class of objects. However, in the 
context of this discussion, it is interesting to consider the top-right panel of Figure 2 in \cite{Michalowski12},
where the stellar masses estimated using standard UV-NIR SED fitting are compared to the 
stellar masses calculated using \grasil modelling of the full UV-NIR SED from \cite{Michalowski10}. Submillimeter 
galaxies correspond to the objects in our analysis where the stellar mass would be most affected by optical-NIR 
attenuation. Therefore, it is striking that in contrast to the results from \cite{LoFaro13}, there does not appear to 
be a significant systematic difference between the stellar masses estimated using the standard UV-NIR and \grasil-based 
UV-FIR methods of fitting submillimeter galaxy SEDs, at least compared to the uncertainties associated with choosing an 
appropriate SFH. 

It is important to note that the SED fitting method applied in \cite{Michalowski10} differs from \cite{LoFaro13} in 
that they use set of template SEDs from \cite{Iglesias07}. These templates were constructed for a limited range of 
the possible parameter space in \grasil, chosen to reproduce the SEDs of star-forming galaxies in the local Universe. 
In contrast, \cite{LoFaro13} do not impose any strong priors on the various free parameters in \grasil, exploring a 
large parameter space. This difference in approach could potentially explain how stellar mass estimation from 
standard UV-NIR SED fitting is only found to be strongly affected by \cite{LoFaro13}. Clearly, for the complex
UV-FIR SED fitting procedures applied by \cite{Michalowski10} and \cite{LoFaro13}, the resultant
stellar masses will depend strongly on the priors and assumptions that are adopted. 

\cite{DaCunha10} fit the UV-FIR SEDs of 16 local ultraluminous infrared galaxies using an alternative procedure to \grasil. Although they do not 
attempt to compare the stellar masses of these objects estimated using their UV-FIR SED fitting method with what would be estimated from standard 
UV-NIR SED fitting, it can be clearly seen that their fitting procedure favours a significant level of dust attenuation at 
optical-NIR wavelengths for all of the objects in their sample. This is consistent with the behaviour revealed 
by our analysis and with \cite{LoFaro13}.

In the local Universe, the problems associated with assuming a specific star-dust geometry can be lessened for resolved, 
low inclination galaxies. \cite{Zibetti09} show that when optical-NIR SED fitting is applied locally to derive the
stellar mass density at each pixel in the images of resolved galaxies with prominent dust lanes, the total stellar mass 
calculated can be higher by up to $40 \, \%$ relative to the stellar mass obtained by fitting integrated photometry.
This result is consistent with the trends shown in our analysis and can be understood as the result of the light emitted
by stars that reside either within or behind optically thick dust lanes being subdominant in the total galaxy SED,
compared to the light emitted by stars on unobscured sightlines. Unfortunately, this method cannot be readily extended
to high redshift where very dusty galaxies are more common.

\subsection{How should theoretical galaxy formation models be compared to observational data?}
\label{DiscussionModelData}

Our analysis is intended to demonstrate that aside from the well documented uncertainties on stellar mass estimation 
associated with SPS modelling and the form of the IMF \cite[e.g.][]{Conroy09}, stellar mass estimation can also be 
significantly affected by the combined effects of dust, metallicity and recycling. However, viewed from another 
perspective our method of applying SED fitting to model galaxy SEDs offers, in principle, a new way to compare 
predictions that involve stellar mass to the results from observational studies. This approach is attractive because 
it allows models to be self-consistently compared with different observational data sets without the need to change 
various parameters in the model (SPS model, IMF) in each instance to make a fair comparison. In addition, the 
scatter between intrinsic and estimated stellar mass is self-consistently accounted for using this method. This 
alleviates the need to invoke an arbitrary level of scatter in order to create agreement between model predictions 
and observations \cite[e.g.][]{Guo11, Bower12}. In the case where predictions from a theoretical model are compared 
simultaneously to both observables and inferred quantities such as stellar mass, it is clear that our methodology 
should be followed to make the comparisons self-consistent with each other. Otherwise, the process of transforming 
from observables to intrinsic galaxy properties will become confused. For example, assumptions made in theoretical 
models to predict luminosity functions are likely to be in conflict with the assumptions made in SED fitting to estimate 
stellar mass functions. Unless our methodology is followed, using both of these diagnostics at the same time could 
therefore adversely affect any attempt to constrain the underlying physics of galaxy formation.

It is important to realise that our methodology does not avoid the problem of converting intrinsic galaxy properties 
into observables. Instead, the burden of accounting for these uncertainties is simply shifted from the observational 
SED fitting procedure back to the theoretical modelling process. The natural alternative is to only consider intrinsic 
galaxy properties and leave all of the uncertainties in the observational process of estimating these quantities.
This approach is widely used in both the semi-analytic modelling and hydrodynamical simulation communities 
\cite[e.g.][]{Bower06, Neistein10, Dave11, Guo11, Khochfar11, Lagos11a, Lamastra13, Ciambur13}. \cite{Conroy10a} 
advocate using this latter approach with the caveat that derived quantities such as stellar mass should only be 
calculated with the inclusion of a full marginalisation over all of the relevant uncertainties in SPS and dust modelling. 
Such a process would require that these uncertainties can be fully characterized by a discrete set of parameters and 
that robust prior distributions for the plausible ranges of these parameters can be found. Our study serves to 
emphasize that choosing an appropriate distribution of priors would be extremely difficult. For example, the extinction 
optical depth of diffuse dust, assumed to be small in  \cite{Conroy10a}, turns out to play a very important role in 
stellar mass estimation for our analysis when the optical depth becomes large.

Finally, Fig.~\ref{mf_evo.Lagos12} and Fig.~\ref{mf_evo.Lacey13} show that the systematics and biases that affect 
stellar mass estimation in our analysis cannot fully account for the level of disagreement between the stellar mass 
functions predicted by \galform models and the mass functions estimated from observational data. Given that these 
models were tuned to reproduce luminosity function data, this demonstrates that stellar mass functions contain 
complementary information to luminosity functions, independent of the uncertainties associated with converting 
intrinsic galaxy properties into observables. This is encouraging and suggests that provided the comparison between 
theoretical models and observations is made self-consistently, existing estimates of the stellar mass function can 
provide significant constraining power for galaxy formation models.

\section{Summary}
\label{ConclusionsSection}

Motivated by the desire to understand whether stellar mass is an appropriate tool for constraining hierarchical galaxy formation models, we have used the observational 
technique of SED fitting to estimate the stellar masses of model galaxies from the semi-analytic model \galform. Following the standard SED fitting procedure for fitting 
broad-band photometry, we find that effects associated with metallicity, recycling and dust can bias stellar mass estimates. In some specific cases, these effects can 
create systematic errors in stellar mass that are comparable to or greater than the potential systematic errors associated with the uncertain form of the IMF.
Furthermore, we have shown that these error sources are often stellar mass dependent, such that the stellar mass function of model galaxies recovered using 
SED fitting can differ substantially in shape as well as in normalization from the intrinsic mass function predicted by a given model.

The cause and nature of the individual systematic error sources uncovered by our analysis are as follows:

\begin{itemize}

\item The exponentially declining star-formation histories that are typically assumed in SED fitting do not, averaged over the entire galaxy population, create
any significant systematic errors in stellar mass. In addition, when averaged over the entire galaxy population, the random errors in stellar mass caused by fitting
with exponentially declining SFHs are small. This is demonstrated in Fig.~\ref{mass.recovery.flythrough.Lagos12}a and Table~\ref{SPSimfRec}. These results are selection 
dependent. If model galaxies are deliberately selected to be undergoing bursts of star formation, the assumption of an exponentially declining SFH leads to both 
systematic underestimates and a significantly larger scatter in the estimated stellar masses of these systems.

\item Differing assumptions regarding recycling of mass from stars back into the ISM can lead to small, redshift dependent systematics in stellar mass. These are outlined
in Table~\ref{SPSimfRec}. Theoretical galaxy formation models typically apply the instantaneous recycling approximation, whereas standard SED fitting procedures use the best-fitting
template SFH to estimate the recycled mass. Neither approach will give the correct answer in detail. Furthermore, the systematic differences between the two approaches
should be accounted for when the stellar masses predicted by theoretical models that assume instantaneous recycling are compared to observational data.

\item Metallicity has the effect of introducing undesirable bimodal features into the distribution of recovered stellar mass that can be seen in Fig~\ref{mass.recovery.flythrough.Lagos12}c. This 
behaviour arises because the standard SED fitting procedure uses discrete, poorly sampled metallicity grids and a statistical method of choosing only a single best-fitting template (the
mode of the probability distribution). 
Alternatively, if the equally common choice of fixing metallicity in SED fitting is implemented, the resultant estimated stellar mass suffers a strong mass-dependent 
bias. These problems can be solved in a straightforward manner by following two simple steps. Firstly, interpolation can be used to fill in the gaps of the original metallicity 
grid provided for publicly available SPS models. Secondly, the statistical technique advocated by \cite{Taylor11} can be implemented where the mean over the probability distribution,
calculated from a likelihood-weighted average over all templates, is used to calculate a best estimate for the stellar mass of a given galaxy.

\item Dust attenuation in massive, dusty galaxies causes standard SED fitting procedures that assume a Calzetti law to systematically underestimate stellar mass. This
occurs because the radiative transfer calculations performed in \galform predict significant dust attenuation at optical-NIR wavelengths in some cases. Thus, the light emitted by
obscured stellar populations in these galaxies is not properly accounted for when estimating stellar mass. Furthermore, either including or excluding any dust attenuation in the SED fitting
process using the Calzetti prescription has only a negligible impact on the estimated stellar mass. This suggests that, for the purposes of stellar mass estimation, it is unimportant whether dust attenuation is included
in the fitting process.

\end{itemize}

We find that the shape of the stellar mass function at $z=0$ is robust against these error sources. However, at higher redshifts the systematic errors associated with dust
significantly reshape the recovered mass functions such that a clear break in the intrinsic model mass function at these redshifts can be blurred out. Furthermore, the effects of
dust can reduce the normalization at the high mass end by up to $0.6 \, \mathrm{dex}$ in some cases.
We are forced to conclude that any attempt to constrain theoretical galaxy formation models using stellar mass functions from high redshift galaxy samples should only be performed
with great care, given the potential for large mass-dependent systematics in stellar mass estimation from SED fitting.

\section*{Acknowledgements}

We thank Danilo Marchesini and Ian Smail for useful discussions.
PDM acknowledges the support of an STFC studentship.
This work was supported in part by an STFC Rolling Grant to the Institute for Computational Cosmology.
Calculations for this paper were performed on the ICC Cosmology Machine, which
is part of the DiRAC Facility jointly funded by STFC, the Large
Facilities Capital Fund of BIS, and Durham University.

\bibliographystyle{mn2e}
\bibliography{bibliography}

\section*{Appendix: stellar mass functions for the Baugh05 model}

\begin{figure*}
\begin{center}
\includegraphics[width=40pc]{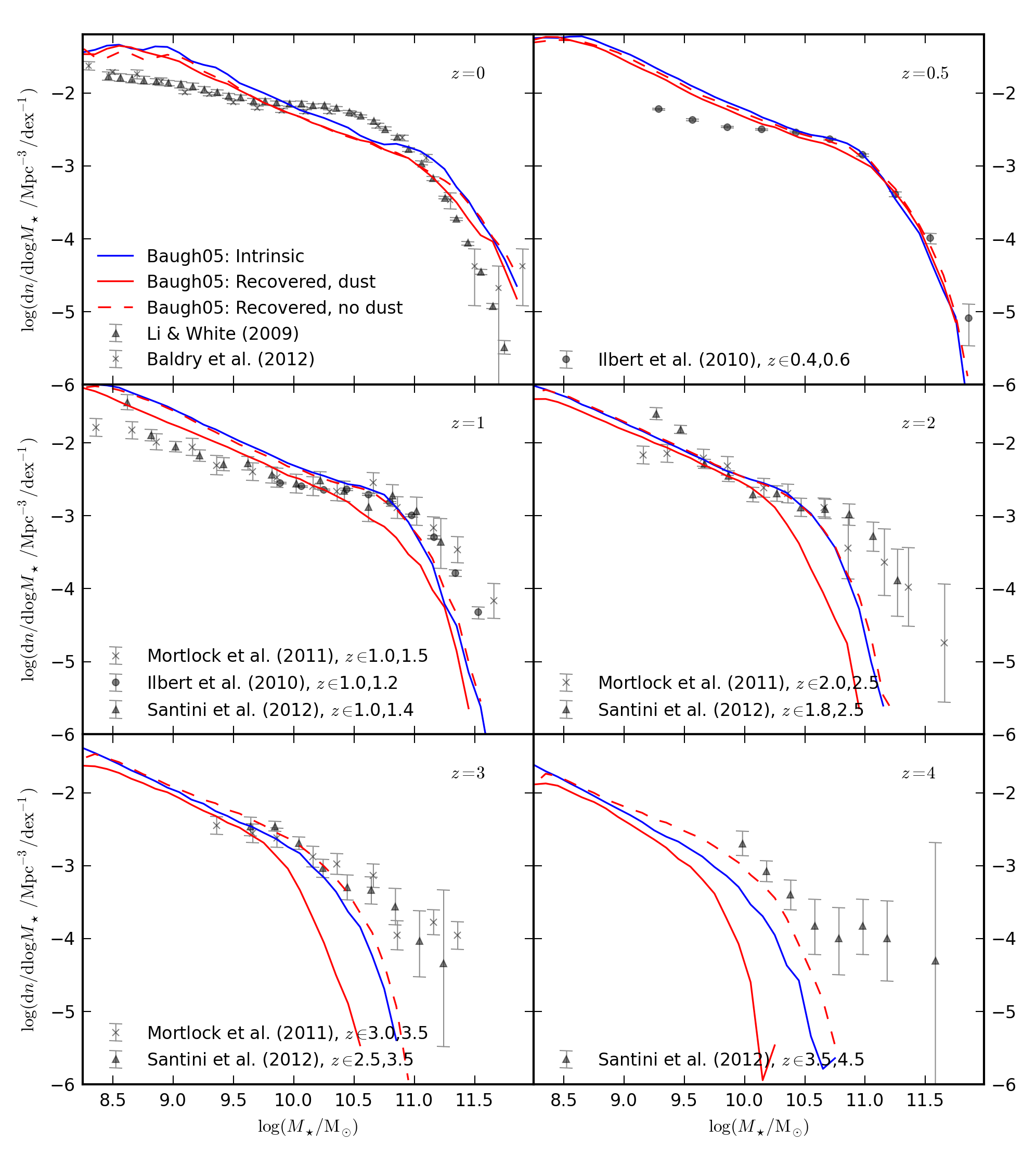}
\caption{Stellar mass functions predicted by the Baugh05 model for a selection of redshifts, as labelled in each panel.
The solid blue line shows the intrinsic stellar mass function produced by the Baugh05 model. 
The solid red line shows the stellar mass function recovered using SED fitting when dust effects are included and a Chabrier IMF is assumed in the fitting procedure. 
As a reference, the dashed red line shows the corresponding stellar mass function where no dust extinction is applied to the model galaxy SEDs and $E(B-V) = 0$ is used as a constraint in the fitting procedure. 
The grey points and error bars show observational estimates of the stellar mass function from \protect \cite{Li09}, \protect \cite{Baldry12}, \protect \cite{Ilbert10}, \protect \cite{Santini12} and \protect \cite{Mortlock11}. 
Where necessary we convert these observational results from a Salpeter to a Chabrier IMF using a $-0.24\,\mathrm{dex}$ correction, calculated by comparing the recovered stellar mass using Salpeter and Chabrier IMFs with
BC03 SPS models.}
\label{mf_evo.Baugh12}
\end{center}
\end{figure*}

Throughout the main text we consider the Lagos12 model from \cite{Lagos12} and the Lacey13 model from Lacey et al. (2013, in preparation). In this appendix, we also
consider the model described in \cite{Baugh05} (hereafter Baugh05). The Baugh05 model is distinct from the Lagos12 and Lacey13 models in that it does not include bursts of star
formation triggered by disk instabilities or the updated star-formation law described in \cite{Lagos11a}. The Baugh05 model also uses different time-scales for star formation, both 
in galaxy disks and in bursts, and uses supernova driven superwinds instead of AGN feedback as a mechanism to suppress the bright end of the luminosity function. Finally, as
described in Section~\ref{ModelDescription}, the Baugh05 model uses a top-heavy IMF in bursts with a slope of $x=0$. This is more extreme than the $x=1$ slope used in the Lacey13 model.

We present stellar mass functions for a selection of redshifts from the Baugh05 model in Fig.~\ref{mf_evo.Baugh12}. Neither the intrinsic or recovered stellar mass functions
agree with the observational estimates of the stellar mass function at $z=0$. The model overpredicts the abundance of low mass galaxies at $z \leq 1$ and overpredicts the abundance
of the most massive galaxies at $z=0$, suggesting that the feedback schemes implemented in this model could be unrealistic. Similar behaviour regarding the effect of dust on 
the recovered stellar mass functions is seen with respect to the Lagos12 and Lacey13 models. At $z=0$, the recovered stellar mass functions (both including and excluding dust attenuation
effects) are lower in normalization with respect to the intrinsic model mass function. This could be a result of the SPS models used in the Baugh05 model. Alternatively, the difference
could be caused by the top-heavy IMF in bursts. We will explore this in more detail in future work.

\label{lastpage}
\end{document}